\documentclass[12pt]{article}
\usepackage{amsmath,amssymb}
\usepackage{bm}
\usepackage{mathrsfs}
\usepackage{fourier}
\usepackage{multirow}
\allowdisplaybreaks[4]
\newcommand{{\Slashp}}{p\!\!\!\!\!\big/}
\newcommand{{\Slashq}}{q\!\!\!\!\!\big/}
\usepackage[usenames]{color}

\begin{document}

\title{Gauge hierarchy problem, supersymmetry,\\
and fermionic symmetry}

\author{
Yoshiharu \textsc{Kawamura}\footnote{E-mail: haru@azusa.shinshu-u.ac.jp}\\
{\it Department of Physics, Shinshu University, }\\
{\it Matsumoto 390-8621, Japan}\\
}

\date{
November 11, 2013}

\maketitle

\begin{abstract}
We reconsider the gauge hierarchy problem
from the viewpoint of effective field theories and a high-energy physics,
motivated by the alternative scenario that the standard model holds 
up to a high-energy scale such as the Planck scale.
The problem is restated as the question
whether it is possible to construct a low-energy effective theory 
and the interaction with heavy particles,
without spoiling the structure of a high-energy physics
supported by an excellent concept.
Based on this reinterpretation,
we give a conjecture that 
theories with hidden fermionic symmetries 
can be free from the gauge hierarchy problem
and become candidates of the physics beyond and/or behind the standard model,
and present toy models with particle-ghost symmetries
as illustrative examples and
a prototype model for the grand unification.
\end{abstract}


\maketitle

\section{Introduction}
\label{Introduction}

Recent experimental results at the Large Hadron Collider (LHC) would revisit
the gauge hierarchy problem~\cite{GH1,GH2},
because the Higgs boson has been found 
with $m_{\rm h} \doteqdot 126$GeV~\cite{LHC1,LHC2}, 
and evidences from new physics 
such as spacetime supersymmtry (SUSY), compositeness and extra dimensions 
have not yet been discovered.

The gauge hierarchy problem is related to the feature that
{\it an effective field theory becomes unnatural,
because fine tuning is required to obtain the weak scale 
and/or to stabilize it against radiative corrections, 
if there is a high-energy physics such as a grand unified theory (GUT) 
relevant to the standard model (SM)}.
It is summarized as the following questions.
What generates the weak scale or the Higgs boson mass?
How is it stabilized?
For example, logarithmic divergences in radiative corrections due to heavy particles
can ruin the stability of the weak scale.

There are at least three possibilities for the problem.
First one is that there is a new physics at the terascale
with a new concept such as SUSY, compositeness
and/or extra dimensions, to solve the problem completely.
Second one is that there is a new physics at the terascale to derive the weak scale,
and the scale is stabilized by some excellent mechanism and/or symmetry 
at a high-energy scale $M_{\rm U}$ such as the Planck scale $M_{\rm Pl}$.
Third one is that there is no new physics concerning the Higgs boson mass at the terascale,
and a high-energy physics solves the problem completely.

In this paper, we reconsider one side of the problem
$\lq\lq${\it how is the weak scale stabilized?}'',
from the viewpoint of effective field theories and a high-energy physics.
Our study is motivated by the alternative scenario 
that the SM (modified with massive neutrinos) 
holds up to $M_{\rm Pl}$~\cite{Shaposhnikov,Nielsen}
and the guiding principle that the gauge hierarchy is stabilized by a symmetry
that should be unbroken in the SM~\cite{Dienes}.
Based on a reinterpretation of the problem from the perspective of a high-energy physics,
we give a conjecture that 
theories with fermionic symmetries different from spacetime SUSY 
can be free from the gauge hierarchy problem
and become candidates of the physics beyond and/or behind the SM,
and present toy models with particle-ghost symmetries
as illustrative examples and
a prototype model to explain the unification of the SM gauge coupling constants,
the triplet-doublet splitting of Higgs boson, and the longevity of proton.

The outline of this paper is as follows.
In the next section, we review the gauge hierarchy problem 
and discuss it in relation to masslessness.
We reconsider the problem from the viewpoint of a high-energy physics,
point out that specific fermionic symmetries 
can play the important role to stabilize a mass hierarchy, 
and propose a grand unification scenario based on the conjecture, in Sect. 3.
In the last section, we give conclusions and discussions.

\section{Gauge hierarchy problem and masslessness}
\label{Gauge hierarchy problem and masslessness}

\subsection{Gauge hierarchy problem}
\label{Gauge hierarchy problem}

We discuss a fine tuning among parameters, 
from the viewpoint of effective field theories.

After subtracting quadratic and quartic divergences if exist,
radiative corrections on parameters $a_i$, up to finite corrections, are given by
\begin{eqnarray}
\delta a_i = \sum_j \frac{c_{ij}}{(4\pi)^2} a_j \ln \frac{\Lambda^2}{\mu^2}~,~~ (i, j = 1, \cdots, n)~,
\label{delta-a}
\end{eqnarray} 
where $c_{ij}$ are functions of parameters,
$\Lambda$ is a cutoff scale, and $\mu$ is a renormalization point.
From the feature that $\delta a_i \to 0$ in the limit of $a_i \to 0$,
the smallness of $a_i$ is understood, if the magnitude of $a_i$ is small enough.
If physical parameters are determined without fine tuning,
the condition $c_{ij} a_j \le O(a_i)$ is roughly imposed on $a_i$.

Fine tuning is, in general, necessary, 
if there is a physics relevant to the SM 
at a higher energy scale beyond the terascale.
For instance, in the presence of heavy particles with masses $M_I$
and some SM gauge quantum numbers,
the radiative corrections on the Higgs mass squared are given by
\begin{eqnarray}
\delta m_{\rm h}^2 = \tilde{c}_{\rm h} \Lambda^2 
+ c'_{\rm h} m_{\rm h}^2 \ln \frac{\Lambda^2}{m_{\rm h}^2}
+ \sum_{I} c''_{{\rm h}I} M_{I}^2 \ln \frac{\Lambda^2}{M_{I}^2} + \cdots~,
\label{deltamhMI}
\end{eqnarray}
where $\tilde{c}_{\rm h}$, $c'_{\rm h}$ and $c''_{{\rm h}I}$ are functions of parameters.
From (\ref{deltamhMI}), we find that the fine tuning is indispensable 
for $c''_{{\rm h}I} M_{I}^2 \gg m_{\rm h}^2$
due to the appearance of the quadratic terms of $M_{I}$ (part of the logarithmic divergences),
even if the quadratic divergence $\tilde{c}_{\rm h} \Lambda^2$ is removed and
unless some miraculous cancellation mechanism works among corrections due to heavy particles.
This induces the technical side of the gauge hierarchy problem~\cite{GH1,GH2}, i.e.,
{\it an unnatural fine tuning is required to stabilize the weak scale
against radiative corrections, 
if there is a high-energy physics such as a GUT relevant to the SM.}

\subsection{Possible solutions}
\label{Possible solutions}

If nature dislikes fine tuning among parameters,
there must exist a reasonable explanation about the absence of fine tuning.
Here, we review some possibilities.\\
~~(1) The concept of the first one is $\lq\lq$compositeness'', i.e., 
some particles are not elementary but composite, made of a more fundamental constituents.
We assume that there is a new dynamics at the terascale, to compose some SM particles.
The typical example is a model that the Higgs doublet is made of new fermions \cite{TC1,Susskind,tHooft}.
The existence of new particles and strong dynamics among them is predicted at the terascale.\\
~~(2) The concept of the second one is $\lq\lq$symmetry'', 
that protects physical parameters against large radiative corrections.
As a new symmetry appearing at the terascale, we enumerate three candidates.\\
~~~(2.1) Supersymmetry~\cite{Veltman,SUSY2}.~~
The SUSY must be realized in a broken phase if exists.
In the presence of soft SUSY breaking terms, $\delta m_{\rm h}^2$ is given by
\begin{eqnarray}
\delta m_{\rm h}^2 = \hat{c}'_{\rm h} m_{\rm h}^2 \ln \frac{\Lambda^2}{m_{\rm h}^2}
+ \hat{c}''_{\rm h} m_{\rm soft}^2 \ln \frac{\Lambda^2}{m_{\rm soft}^2} + \cdots~,
\label{deltamh-soft}
\end{eqnarray} 
where $\hat{c}'_{\rm h}$ and $\hat{c}''_{\rm h}$ are functions of parameters,
and $m_{\rm soft}$ is a typical mass parameter representing the soft SUSY breaking.
The magnitude of $m_{\rm soft}$ is a same order of masses of superpartners for the SM particles.
The absence of fine tuning requires roughly $m_{\rm soft} \le O(1)$TeV, 
and then the existence of superpartners concerning the SM particles is predicted at the terascale.\\
~~~(2.2) Global symmetry.~~
The Higgs boson can be a pseudo Nambu-Goldstone boson
relating a spontaneous breakdown of a global symmetry~\cite{IK&T}.
The smallness and the stability of the Higgs boson mass 
and the smallness of Yukawa coupling constants 
arise from the nature of Nambu-Goldstone particle.\\
~~~(2.3) Conformal symmetry~\cite{Frampton}.~~
The quantum conformal invariance in collaboration with finiteness,
which is called $\lq\lq$conformality'', can solve the problem.
In the appearance of new particles at the terascale,
the theory becomes scale invariant with the vanishing $\beta$ functions.
Then, physical parameters do not run beyond the scale, and
the concept of scale becomes vague.\\
~~(3) The concept of the third one is $\lq\lq$extra dimensions'', i.e.,
there exists extra spatial dimensions other than 4-dimensional spacetime.
We assume that there is a fundamental theory at the terascale with a fundamental mass 
parameter of $O(1)$TeV, concerning extra dimensions.
The typical examples are models with large extra dimensions~\cite{ED1,ED2}.

The combination of $\lq\lq$extra dimensions'' and $\lq\lq$symmetry''
produces a new solution, which is called $\lq\lq$gauge-higgs unification''~\cite{GHU1,GHU2}.
The extra-dimensional component $(A_y)$ of gauge field is massless at the tree level
due to the gauge invariance,
and receives a finite correction on its mass upon compactification~\cite{HIL}.
By the identification of $A_y$ with the Higgs boson, $m_{\rm h}$ becomes a natural parameter
because gauge symmetry enhances on the higher-dimensional one in the limit of $m_{\rm h} \to 0$,
and the weak scale is stabilized 
in the case with a large compactification scale of $O(1)$TeV$^{-1}$.

The stabilization of the extra-dimensional space is crucial 
for the solution to the gauge hierarchy problem
in theories on a higher-dimensional space-time,
including the Randall-Sundrum model~\cite{R&S}.\\
~~(4) There is a new physics at a higher energy scale $M_I$
than the terascale, but the interaction with the SM is extremely weak.
Or large radiative corrections are not induced,
if the mixing among parameters of the physics
at $M_I$ and the SM is tiny enough such that $c''_{{\rm h}I} \le O(m_{\rm h}^2/M_I^2)$.\\
~~(5) The SM (or an extension of the SM with new particles around the terascale)
holds up to a high energy scale $M_{\rm U}$,
without a new concept to stabilize the Higgs boson mass at the terascale.
We assume that a new physics appears at $M_{\rm U}$,
which is described by an ultimate theory,
the initial value of $m_{\rm h}$ is fixed by the new physics,  
and some mechanism and/or symmetry protects $m_{\rm h}^2$ against large radiative corrections~\cite{Dienes,A&I,K}.

There is a possibility that a new physics and/or concept is hidden behind the SM, too.

\subsection{Masslessness and finiteness}
\label{Masslessness and finiteness}

Before we reexamine the gauge hierarchy problem from a different angle,
we discuss its related topics on a basis of an ultimate theory. 

First, we assume that the physics at $M_{\rm U}$ is described by an ultimate theory,
which has $\lq\lq$finiteness'', i.e., physical quantities are calculated as finite values.
At a rough guess, the magnitude of quantities 
with mass dimension $d$ is estimated as $O(M_{\rm U}^d)$,
and natural initial conditions for masses of particles in low-energy physics
would be given by
\begin{eqnarray}
m_i(M_{\rm U}) = 0~,
\label{m(M)}
\end{eqnarray}
as far as a mechanism to generate a tiny value does not work in the ultimate theory.
We refer to the relations $m_i(M_{\rm U}) = 0$ as $\lq\lq$masslessness''.\footnote{
Masslessness could be related to the vanishment of bare Higgs boson mass around $M_{\rm Pl}$~\cite{HK&O}.
}
Non-zero masses and scales are expected to be dynamically generated 
by quantum effects, in the effective field theory.
In other words,
it might be a natural choice that every particle in a low-energy theory is massless at $M_{\rm U}$,
the effective theory has the classical conformal symmetry in addition to
chiral symmetry and gauge symmetry, and masses are induced after the breakdown of relevant
symmetries by some dynamics.
The typical examples are the Higgs mechanism in electroweak theory
and the dimensional transmutation in quantum chromodynamics.

At this stage, the following questions arise.
{\it What is the origin of masslessness and finiteness?
How is masslessness protected against quantum effects?}

We need to specify an ultimate theory, in order to answer the above questions.
Here, we take string theory as a possible candidate.\footnote{
As a candidate of field theory version,
the theories called finite unified theories, which have a large predictive power,
have been proposed \cite{FUT}.
They are based on the finiteness and the principle of reduction of coupling constants.
}
In string theory, the world-sheet conformal invariance induces the massless string states,
and the world-sheet modular invariance guarantees finiteness of physical quantities.
Hence, it can be said that the world-sheet modular invariance 
is responsible for the protection of masslessness against quantum corrections.

Concretely, from the world-sheet modular invariance for the closed string,
the correction $\delta m_{\phi}^2$ (radiative corrections of the scalar mass squared 
including contributions from innumerable string states)
should be given by
\begin{eqnarray}
\delta m_{\phi}^2 =
\int_{\mathcal{F}} \frac{d^2\tau}{\tau_2^2} G(\tau)~,
\label{deltamphi-st}
\end{eqnarray}
where $\tau = \tau_1 + i \tau_2$ is a modular parameter,
$G(\tau)$ is a worldsheet modular invariant function, i.e., $G(\tau) = G(\tau+1)$
and $G(\tau) = G(-1/\tau)$,
and $\mathcal{F}$ stands for the fundamental region defined by
$\mathcal{F} = \{\tau: |\mbox{Re}\tau| \le {1}/{2}, 1 \le |\tau|\}$.
In cases where SUSY holds exactly, $G(\tau)$ vanishes, and then $\delta m_{\phi}^2 = 0$.
Even if SUSY is broken down, there is a possibility that $G(\tau)$ vanishes in conspiracy with
infinite towers of massive particles,
as suggested in Ref.~\cite{Dienes}.

From the viewpoint of effective field theories, symmetries relevant to
naturalness such as chiral symmetry, gauge symmetry and conformal symmetry
become useful tools for realistic model-building, that is,
naturalness becomes a powerful guiding principle 
to construct an effective theory~\cite{tHooft}.
The relation of naturalness and conformal symmetry 
has been reexamined by Bardeen~\cite{Bardeen}.\footnote{
Extensions of the SM have been proposed
by adopting the classical conformal invariance as a guiding principle~\cite{M&N,FKM&V,H&K,IO&O,I&O,EJK&S,HRRS&T,H&S,C&R}.
}

On the other hand, from the viewpoint of an ultimate theory,
masslessness is more essential than naturalness or symmetries that make parameters natural.
In other words, naturalness or the relevant symmetry 
is regarded as a secondary concept, originated from masslessness.
Also, there is a possibility that an ultimate theory provides constraints on its effective theory.
For instance, the ultimate theory possesses a duality like the world-sheet modular invariance.
If this symmetry or its remnant could be imposed on the effective theory,
only logarithmic divergent parts might be picked out
and the Higgs boson mass could become a natural parameter,
as discussed in \cite{K2}.

For the gauge hierarchy problem,
the fine tuning of order $(m_{\phi}/M_{\rm U})^2$ is required
for the scalar mass squared $m_{\phi}^2 (\ll M_{\rm U}^2)$
from the viewpoint of effective field theories.
In the ultimate theory, there must be symmetries 
such as world-sheet conformal symmetry, modular invariance and SUSY in string theory,
to generate masslessness and protect it against radiative corrections of $O(M_{\rm U}^2)$.
Hence, we expect that 
such fine tuning might also be an artifact in its effective field theory,
and it could be improved if features of the ultimate theory
are taken in and the ingredients of the effective theory are enriched.

\section{Supersymmetry and fermionic symmetry}
\label{Supersymmetry and fermionic symmetry}

Let us reconsider the technical side of the gauge hierarchy problem, based on the last possibility 
presented in Sect. \ref{Possible solutions},
because it is plausible on the basis of recent experimental results at LHC.
Evidences from SUSY, compositeness and extra dimensions have not yet been discovered,
and a definite discrepancy has not yet been observed 
between the predictions in the SM (modified with massive neutrinos) and experimental results.
These suggest that, even if new particles and/or new dynamics exist, 
those effects must be adequately suppressed.

Our consideration is based on the following assumptions, 
relating a physics beyond and/or behind the SM.\\
(a) There is an ultimate theory at a high-energy scale $M_{\rm U}$,
and it contains particles with masses of $O(M_{\rm U})$ and massless ones.
The physical sector of massless particles is described by 
the SM (or an extension of the SM with new particles around the terascale).
We denote it by SM + $\alpha$.
This model holds up to $M_{\rm U}$, 
and the Higgs boson is described as an elementary particle.\\
(b) There exists a new physics with a new concept $\mathcal{X}$ 
beyond and/or behind SM + $\alpha$,
which is one of characteristics in the ultimate theory.
The new physics can be formulated by an effective field theory possessing $\mathcal{X}$.\\
(c) The full effective theory consists of three parts, 
the part $X_{\rm heavy}$ describing heavy particles
with masses of $O(M_{\rm U})$,
the part $X_{\rm light}$ ($\supset {\rm SM} + \alpha$) 
including light (or massless at $M_{\rm U}$) particles
that survive at a lower-energy scale,
and the part $X_{\rm mix}$ 
describing the interactions between particles in $X_{\rm high}$ and those in $X_{\rm light}$.
The gauge hierarchy problem does not occur, 
and the physical sector can be described by an effective theory 
without $X_{\rm heavy}$ and $X_{\rm mix}$.
This feature consists of the following ingredients.\\
(c1) The physical parameters in SM + $\alpha$ do not receive large quantum corrections,
in the presence of $X_{\rm heavy}$ and $X_{\rm mix}$.\\
(c2) The concept $\mathcal{X}$ is preserved in the full effective theory, 
independent of the behavior of the particles in SM + $\alpha$.

In the following, we search for $\mathcal{X}$ to realize (a) -- (c),
and explore theories with $\mathcal{X}$ beyond and/or behind SM + $\alpha$.

\subsection{Fragility of supersymmetry and gauge hierarchy problem}
\label{Fragility of supersymmetry and gauge hierarchy problem}

For the sake of completeness, we examine whether spacetime SUSY 
is suitable as $\mathcal{X}$ or not,
by making clear the strong and weak points of SUSY, using a toy model.

Let us begin with the Lagrangian density given by
\begin{eqnarray}
&~& \mathcal{L}_{\rm SUSY} = \mathcal{L}_{(0)} + \mathcal{L}_{(1,2)} + \mathcal{L}_{\rm mix}~,
\label{L-SUSY}\\
&~& \mathcal{L}_{(0)} 
= \left|\partial_\mu \phi_0\right|^2 + \overline{\psi}_0 i \overline{\sigma}^{\mu} \partial_{\mu} \psi_0~,
\label{L0}\\
&~& \mathcal{L}_{(1,2)} = \left|\partial_\mu \phi_1\right|^2 - M^2 \left|\phi_1\right|^2
+ \left|\partial_\mu \phi_2\right|^2 - M^2 \left|\phi_2\right|^2 
\nonumber \\
&~& ~~~~~~~~~~~~
+ \overline{\psi}_1 i \overline{\sigma}^{\mu} \partial_{\mu} \psi_1 
+ \overline{\psi}_2 i \overline{\sigma}^{\mu} \partial_{\mu} \psi_2
- M \psi_1 \psi_2 - M \overline{\psi}_1 \overline{\psi}_2~,
\label{L12}\\
&~& 
\mathcal{L}_{\rm mix} = - f^2 \left|\phi_0\right|^2 \left(\left|\phi_1\right|^2 + \left|\phi_2\right|^2\right)
- f^2 \left|\phi_1\right|^2 \left|\phi_2\right|^2
- f M \left(\phi_0 + \phi^*_0\right) \left(\left|\phi_1\right|^2 + \left|\phi_2\right|^2\right)
\nonumber \\
&~& ~~~~~~~~~~~
- f \phi_0 \psi_1 \psi_2 - f \psi_0 \phi_1 \psi_2  - f \psi_0 \psi_1 \phi_2 
- f \phi^*_0 \overline{\psi}_1 \overline{\psi}_2
- f \overline{\psi}_0 \phi^*_1 \overline{\psi}_2 
- f \overline{\psi}_0 \overline{\psi}_1 \phi^*_2 ~,
\label{Lmix}
\end{eqnarray}
where $\phi_k$ and $\psi_k$ $(k=0,1,2)$ are complex scalar bosons and Weyl fermions, respectively,
and parameters $M$ and $f$ are chosen as real, for simplicity.
The SUSY invariance of $\mathcal{L}_{\rm SUSY}$ is understood
from the rewritten version,
\begin{eqnarray}
\mathcal{L}_{\rm SUSY} = \sum_k \left(\left|\partial_\mu \phi_k\right|^2
+ \overline{\psi}_k i \overline{\sigma}^{\mu} \partial_{\mu} \psi_k 
- \left|\frac{\partial W}{\partial \phi_k}\right|^2\right)
- \left(\frac{1}{2} \sum_{k,l} \frac{\partial^2 W}{\partial \phi_k \partial \phi_l} \psi_k \psi_l
 + {\rm h.c.}\right)~,
\label{L-SUSY2}
\end{eqnarray} 
where $W= M \phi_1 \phi_2 + f \phi_0 \phi_1 \phi_2$ and h.c. represents the hermitian conjugate.

From (\ref{L0}) and (\ref{L12}),
we find that, at the tree level, 
both $\phi_0$ and $\psi_0$ are massless,
both $\phi_1$ and $\phi_2$ have a mass $M$, and 
$\psi_1$ and $\psi_2$ form a Dirac fermion with a mass $M$.

Let us suppose that the sectors described by $\mathcal{L}_{(0)}$, $\mathcal{L}_{(1,2)}$ 
and $\mathcal{L}_{\rm mix}$ are counterparts to 
a SUSY extension of SM + $\alpha$, $X_{\rm heavy}$ and $X_{\rm mix}$, respectively.
Although the second term in $\mathcal{L}_{\rm mix}$ contains only heavy fields,
we assume that it belongs to $X_{\rm mix}$ 
because it originates from the mixing of light and heavy fields in $W$.

The non-renormalization theorem states 
that both $M$ and $f$ do not receive any radiative corrections perturbatively, 
and hence the mass spectrum remains unchanged 
and the hierarchical structure holds at the quantum level.
This is the strong point of SUSY, and
SUSY extensions of GUTs become candidates of a theory with $\mathcal{X}$~\cite{Sakai,D&G}.

However, SUSY has not yet been found in particle physics.
Hence, if SUSY exists in nature, it is realized in a broken form.
There are at least two possibilities to explain the current status.
One is that superpartners of the SM particles exist, 
but they are too heavy to observe through the present collider experiments.
Then, naturalness of $m_{\rm h}$ would be viewed with suspicion 
because of the necessity for a (mild) fine tuning, 
as superpartners become heavier~\cite{EEN&Z,B&G}.
The other one is that there are no superpartners of the SM particles at all.
Then, a weakness of SUSY would become apparent (as will be shown below)
because of the missing of SUSY in the low-energy physics 
after the reduction of SUSY and the elimination of superpartners, 
even if SUSY exists at an ultimate theory.

In the following, we hit a sensitive point of SUSY by answering the question
what happens if superpartners are missing, 
considering the case that $\psi_0$ is eliminated in the above toy model.
Or we show that the parameters receive radiative corrections in the absence of $\psi_0$,
and the mass hierarchy is spoiled.
Concretely,
at the one-loop level, the mass squared of $\phi_0$, $m_{\phi_0}^2$,
does not receive any radiative corrections. 
This is due to the fact that $\phi_0$ couples with heavy fields in a SUSY invariant form.
The mass squareds of $\phi_1$ and $\phi_2$
receive the radiative corrections of $O(M^2)$,
even if quadratic divergences are removed.
On the other hand, $\psi_1$ and $\psi_2$ do not receive mass corrections at the one-loop level,
and the mass degeneracy of heavy fields is lost.
Hence $m_{\phi_0}^2$
receives large radiative corrections of $O(M^2)$ up to some suppression factors
at the two-loop level,
and the mass hierarchy is destroyed.

Although we have considered the toy model 
where the system with $\mathcal{L}'_{(0)}=\left|\partial_\mu \phi_0\right|^2$
corresponds to SM + $\alpha$,
it is easily understood that a similar thing happens 
in any SUSY extensions of the SM
and the gauge hierarchy problem occurs.\footnote{
In SUSY extensions of the SM,
quadratic divergences appear in radiative corrections on scalar masses
in the absence of (part of) superpartners, even if we neglect the contributions from heavy particles.
To avoid such a complication and extract effects of heavy particles, 
the toy model is considered. 
}
In this way, SUSY makes a strong showing in the presence of superpartners,
but it is fragile if missing. 

We rethink what happened, from the viewpoint of heavy particles.
Let light fields introduce without their superpartners
into a SUSY invariant system including heavy fields.
The structure of SUSY invariant system is broken down through the coupling to the system
without superpartners, 
and it induces large radiative corrections on masses of light scalar fields.
This is a root or might be an essence of the gauge hierarchy problem.
In other words, the gauge hierarchy problem can be restated as  
$\lq\lq${\it without upsetting the structure of a high-energy physics 
under cover of an excellent concept,
is it possible to formulate a low-energy effective theory 
and the interaction with heavy particles?}"

\subsection{Fermionic symmetry with a charmed life}

First, we speculate a theory whose structure is stabilized by some symmetry $\mathcal{X}$.
The strong point of SUSY provides a useful hint.
It is that the cancellation on radiative corrections 
works very well due to contributions from particles with different statistics,
that form supermultiplets, if SUSY holds exactly.

The spacetime SUSY pairs every particle with its superpartner
whose helicity is one-half different from,
because SUSY charges $(Q_{\alpha}, \overline{Q}_{\dot{\alpha}})$ have helicity $\pm 1/2$
and satisfy the relation 
$\{Q_{\alpha}, \overline{Q}_{\dot{\alpha}}\} = 2 \sigma_{\alpha\dot{\alpha}}^{\mu} P_{\mu}$.
Here, we consider $N=1$ SUSY for simplicity.
Note that $Q_{\alpha}$ and $\overline{Q}_{\dot{\alpha}}$-singlets satisfying
$Q_{\alpha} \psi(x) = 0$ and  $\overline{Q}_{\dot{\alpha}} \psi(x) = 0$ are not allowed 
because of $P_{\mu} \psi(x) = i \partial_{\mu} \psi(x) \ne 0$.
This fact leads to a weak point of SUSY
in case that some superpartners are missing.

From this observation, we anticipate that something quite interesting can happen, 
if there is a symmetry
such that the SM particles are singlets and heavy particles form doublets 
under the transformation group,
and quantum corrections from each component in the doublet are canceled out each other.
Then, a possible candidate of $\mathcal{X}$ is a fermionic symmetry 
that transforms an ordinary particle into its ghost partner.
Here, an ordinary particle means a particle that obeys the spin-statistics theorem.
The ghost partner has same spacetime and internal quantum numbers
with the corresponding ordinary particle, but yields a different statistics.
We refer to this type of fermionic symmetry as a $\lq\lq$particle-ghost symmetry''.

We explore theories with ghost fields, in the expectation that 
the particle-ghost symmetry can play the vital role
to stabilize the Higgs boson mass, although it is hidden behind the SM.

Let us consider a toy model with a complex scalar particle $\phi$ with a light mass $m_{\phi}$
and a pair of complex scalar particles $(\varphi, c_{\varphi})$ with a heavy mass $M_{\varphi}$.
Here, $\varphi$ is an ordinary scalar field yielding the commutation relations
and $c_{\varphi}$ is its ghost partner yielding the anti-commutation relations.
If both $\varphi$ and $c_{\varphi}$ interacts with $\phi$ in the same way,
radiative corrections on $m_{\phi}$ due to heavy fields would vanish 
because of the cancellation between contributions from $\varphi$ and $c_{\varphi}$.
Note that the extra minus sign appears for the virtual ghost field running in the loop.
Furthermore, the mass of $c_{\varphi}$ would receive the same size of radiative corrections
with that of $\varphi$ through the interactions with $\phi$.
Hence we expect that the particle-ghost symmetry is unbroken at the quantum level,
and the mass hierarchy is stabilized.

Next, we embody our speculation, using the Lagrangian density,
\begin{eqnarray}
&~& \mathcal{L}_{\rm T} 
= \mathcal{L}_{\phi} + \mathcal{L}_{\varphi, c} + \mathcal{L}_{\rm mix}~,
\label{L-T}\\
&~& \mathcal{L}_{\phi} = \partial_\mu \phi^{\dagger} \partial^{\mu} \phi 
- m_{\phi}^2 \phi^{\dagger}\phi
- \lambda_{\phi} \left(\phi^{\dagger}\phi\right)^2~,
\label{L-phi}\\
&~& \mathcal{L}_{\varphi, c} 
= \partial_\mu \varphi^{\dagger} \partial^{\mu} \varphi 
+ \partial_\mu c_{\varphi}^{\dagger} \partial^{\mu} c_\varphi
- M_{\varphi}^2 \left(\varphi^{\dagger} \varphi + c_{\varphi}^{\dagger} c_\varphi\right)
\nonumber \\
&~& ~~~~~~~~~~~~~~ 
- \lambda_{\varphi} \left(\varphi^{\dagger} \varphi + c_{\varphi}^{\dagger} c_\varphi\right)
\star \left(\varphi^{\dagger} \varphi + c_{\varphi}^{\dagger} c_\varphi\right)~,
\label{L-varpsi}\\
&~& \mathcal{L}_{\rm mix}
= - \lambda' \phi^{\dagger} \phi \left(\varphi^{\dagger} \varphi + c_{\varphi}^{\dagger} c_\varphi\right)~,
\label{L-mix-varpsi}
\end{eqnarray}
where $\lambda_{\phi}$ is the quartic self-coupling constant of $\phi$,
$\lambda_{\varphi}$ and $\lambda'$ are other quartic coupling constants,
and the star product ($\star$) represents a non-local interaction.
The self-interactions of heavy fields are indispensable,
because they are induced radiatively via the couplings between light and heavy fields.
Features of interaction terms are given in the Appendix A.
The sectors described by $\mathcal{L}_{\phi}$, $\mathcal{L}_{\varphi, c}$
and $\mathcal{L}_{\rm mix}$ are counterparts to SM + $\alpha$,
$X_{\rm heavy}$ and $X_{\rm mix}$, respectively.

Here, we outline radiative corrections on parameters.
Details are presented in the Appendix A.
At the one-loop level, the mass squared of $\phi$ does not receive any radiative corrections
from heavy fields, because the contributions from $\varphi$ and $c_{\varphi}$ 
are exactly canceled out.
On the other hand, the parameters $M_{\varphi}$, $\lambda_{\varphi}$ and $\lambda'$ 
receive radiative corrections through $\mathcal{L}_{\rm mix}$ and the interactions in $\mathcal{L}_{\varphi, c}$.
If both $\varphi$ and $c_{\varphi}$ receive exactly the same size of contributions,
the structure of $\mathcal{L}_{\varphi, c}$ and $\mathcal{L}_{\rm mix}$ remain unchanged.
This can be shown at the one-loop level if interaction terms satisfy some features.
If the stabilization of $\mathcal{L}_{\varphi, c}$ and $\mathcal{L}_{\rm mix}$ 
hold at the all order of perturbation and
the quadratic divergence in the mass squared of $\phi$,
originated from the self-interaction of $\phi$, 
is subtracted, the system is free from fine tuning.
Then, the mass hierarchy can be stabilized
against quantum corrections.

Now, we study a characteristics $\mathcal{X}$ to stabilize the theory.
From (\ref{L-varpsi}) and (\ref{L-mix-varpsi}),
we guess that the quardratic form $\mathcal{I}= \varphi^{\dagger} \varphi + c_{\varphi}^{\dagger} c_\varphi$
is a key and a symmetry relating transformations 
which leave $\mathcal{I}$ invariant can be $\mathcal{X}$.
It is equivalent to $OSp(2|2)$.\footnote{
The $OSp(2|2)$ is the group whose elements are generators of transformations
(corresponding (a), (b) and (d))
which leave the inner product of two vectors such as 
$x_1 x_2 + y_1 y_2 + (\overline{\theta}_1 {\theta}_2 - {\theta}_1 \overline{\theta}_2)/2$
invariant, where $x_i$ and $y_i$ $(i=1, 2)$ 
are bosonic variables, and $\theta_i$ and $\overline{\theta}_i$ are fermionic ones.
Note that the innner product is given by 
$x^2 + y^2 + \overline{\theta} \theta (= |z|^2 + \overline{\theta} \theta)$ for a same vector, 
where $z=x+iy$.
}

The transformations are classified into following four types.\\
(a) $U(1)$ transformation for a particle:
\begin{eqnarray}
\delta_{\rm o} \varphi = i \epsilon_{\rm o} \varphi~,~~ 
\delta_{\rm o} \varphi^{\dagger} = -i \epsilon_{\rm o} \varphi^{\dagger}~,~~
\delta_{\rm o} c_{\varphi} = 0~,~~ \delta_{\rm o} c_{\varphi}^{\dagger} = 0~,
\label{delta-o}
\end{eqnarray}
where $\epsilon_{\rm o}$ is an infinitesimal real number.\\
(b) $U(1)$ transformation for a ghost:
\begin{eqnarray}
\delta_{\rm g} \varphi = 0~,~~ 
\delta_{\rm g} \varphi^{\dagger} = 0~,~~
\delta_{\rm g} c_{\varphi} = i \epsilon_{\rm g} c_{\varphi}~,~~ 
\delta_{\rm g} c_{\varphi}^{\dagger} = -i \epsilon_{\rm g} c_{\varphi}^{\dagger}~,
\label{delta-g}
\end{eqnarray}
where $\epsilon_{\rm g}$ is an infinitesimal real number.\\
(c) Transformation that $c_{\varphi}$ changes into $c_{\varphi}^{\dagger}$ 
and its hermitian conjugation:
\begin{eqnarray}
&~& \delta_{\rm c} \varphi = 0~,~~ 
\delta_{\rm c} \varphi^{\dagger} = 0~,~~
\delta_{\rm c} c_{\varphi} = \epsilon_{\rm c} c_{\varphi}^{\dagger}~,~~ 
\delta_{\rm c} c_{\varphi}^{\dagger} = 0~,
\label{delta-c}\\
&~& \delta_{\rm c}^{\dagger} \varphi = 0~,~~ 
\delta_{\rm c}^{\dagger} \varphi^{\dagger} = 0~,~~
\delta_{\rm c}^{\dagger} c_{\varphi} = 0~,~~ 
\delta_{\rm c}^{\dagger} c_{\varphi}^{\dagger} = \epsilon_{\rm c}^{\dagger} c_{\varphi}~,
\label{delta-cdagger}
\end{eqnarray}
where $\epsilon_{\rm c}$ and $\epsilon_{\rm c}^{\dagger}$
are Grassman numbers.\\
(d) Fermionic transformations (particle-ghost transformations):
\begin{eqnarray}
&~& \delta_{\rm F} \varphi = -\zeta c_{\varphi}~,~~\delta_{\rm F} \varphi^{\dagger} = 0~,~~ 
\delta_{\rm F} c_{\varphi} = 0~,~~
\delta_{\rm F} c_{\varphi}^{\dagger} = \zeta \varphi^{\dagger}~,
\label{delta-F} \\
&~& \delta_{\rm F}^{\dagger} \varphi = 0~,~~
\delta_{\rm F}^{\dagger} \varphi^{\dagger} = \zeta^{\dagger} c_{\varphi}^{\dagger}~,~~
\delta_{\rm F}^{\dagger} c_{\varphi} = \zeta^{\dagger} \varphi~,~~
\delta_{\rm F}^{\dagger} c_{\varphi}^{\dagger} = 0~,
\label{delta-Fdagger}
\end{eqnarray}
where $\zeta$ and $\zeta^{\dagger}$ are Grassman numbers.
Note that $\delta_{\rm F}$ is not generated by a hermitian operator, 
different from the generator of the BRST transformation
in systems with first class constraints~\cite{BRST} 
and that of the topological symmetry~\cite{W,Top}.

From the above transformation properties, 
we see that ${\bm \delta}_{\rm c}$, ${\bm \delta}_{\rm c}^{\dagger}$,
${\bm \delta}_{\rm F}$ and ${\bm \delta}_{\rm F}^{\dagger}$ are nilpotent, i.e.,
${\bm \delta}_{\rm c}^2 = 0$, ${\bm \delta}_{\rm c}^{\dagger2}=0$, ${\bm \delta}_{\rm F}^2= 0$
and ${\bm \delta}_{\rm F}^{\dagger2}=0$,
where ${\bm \delta}_{\rm c}$, ${\bm \delta}_{\rm c}^{\dagger}$, 
${\bm \delta}_{\rm F}$ and ${\bm \delta}_{\rm F}^{\dagger}$ are 
defined by $\delta_{\rm c} = \epsilon_{\rm c} {\bm \delta}_{\rm c}$, 
$\delta_{\rm c}^{\dagger} = \epsilon_{\rm c}^{\dagger} {\bm \delta}_{\rm c}^{\dagger}$
$\delta_{\rm F} = \zeta {\bm \delta}_{\rm F}$ 
and $\delta_{\rm F}^{\dagger} = \zeta^{\dagger} {\bm \delta}_{\rm F}^{\dagger}$, respectively.
Furthermore, we find the algebraic relations,
\begin{eqnarray}
\left\{Q_{\rm c}, Q_{\rm c}^{\dagger}\right\} = N_{\rm g}~,~~
\left\{Q_{\rm F}, Q_{\rm F}^{\dagger}\right\} = N_{\rm o} + N_{\rm g} \equiv N_{\rm D}~,
\label{QQdagger}
\end{eqnarray}
where $N_{\rm o}$, $N_{\rm g}$, $Q_{\rm c}$, $Q_{\rm c}^{\dagger}$, $Q_{\rm F}$ and $Q_{\rm F}^{\dagger}$ are the corresponding generators (charges)
given by
\begin{eqnarray}
&~& \delta_{\rm o} \Phi = i\left[\epsilon_{\rm o} N_{\rm o}, \Phi\right]~,~~
\delta_{\rm g} \Phi = i\left[\epsilon_{\rm g} N_{\rm g}, \Phi\right]~,~~
\delta_{\rm c} \Phi = i\left[\epsilon_{\rm c} Q_{\rm c}, \Phi\right]~,~~
\nonumber \\
&~& \delta_{\rm c}^{\dagger} \Phi 
= i\left[\epsilon_{\rm c}^{\dagger} Q_{\rm c}^{\dagger}, \Phi\right]~,~~
\delta_{\rm F} \Phi = i\left[\zeta Q_{\rm F}, \Phi\right]~,~~
\delta_{\rm F}^{\dagger} \Phi = i\left[Q_{\rm F}^{\dagger}\zeta^{\dagger}, \Phi\right]~.
\label{Qs}
\end{eqnarray}

It is easily understood that $\varphi^{\dagger} \varphi + c_{\varphi}^{\dagger} c_\varphi$ is invariant 
under the transformations (\ref{delta-F}) and (\ref{delta-Fdagger}),
from the nilpotency of ${\bm \delta}_{\rm F}$ and ${\bm \delta}_{\rm F}^{\dagger}$ and the relations,
\begin{eqnarray}
&~& \varphi^{\dagger} \varphi + c_{\varphi}^{\dagger} c_\varphi
= {\bm \delta}_{\rm F}\left(c_{\varphi}^{\dagger} \varphi\right)
= {\bm \delta}_{\rm F}^{\dagger}\left(\varphi^{\dagger} c_\varphi\right)
= {\bm \delta}_{\rm F}  {\bm \delta}_{\rm F}^{\dagger} \left(\varphi^{\dagger} \varphi\right) 
= -{\bm \delta}_{\rm F}^{\dagger}  {\bm \delta}_{\rm F}
\left(\varphi^{\dagger} \varphi\right)
\nonumber \\
&~& ~~~~~~~~~~~~~~~~~~~~~~
= -{\bm \delta}_{\rm F}  {\bm \delta}_{\rm F}^{\dagger} 
\left(c_{\varphi}^{\dagger} c_{\varphi}\right) 
= {\bm \delta}_{\rm F}^{\dagger}  {\bm \delta}_{\rm F}
\left(c_{\varphi}^{\dagger} c_{\varphi}\right)~.
\label{delta-rel}
\end{eqnarray}
Using them, the Lagrangian density $\mathcal{L}_{\rm T}$ is rewritten as
\begin{eqnarray}
\mathcal{L}_{\rm T} 
= \mathcal{L}_{\phi} 
+  {\bm \delta}_{\rm F}  {\bm \delta}_{\rm F}^{\dagger} 
\left[\partial_\mu \varphi^{\dagger} \partial^{\mu} \varphi 
- M_{\varphi}^2 \varphi^{\dagger} \varphi 
- \frac{\lambda_{\varphi}}{2} \left(\varphi^{\dagger} \varphi \star \varphi^{\dagger} \varphi 
- c_{\varphi}^{\dagger} c_\varphi \star c_{\varphi}^{\dagger} c_\varphi\right)
- \lambda' \phi^{\dagger} \phi \varphi^{\dagger} \varphi\right]~.
\label{L-T-again}
\end{eqnarray}

The theory is specified by the fermionic charges  $Q_{\rm F}$ and $Q_{\rm F}^{\dagger}$
and the bosonic charge $N_{\rm D}$ of the doublet $(\varphi, c_{\varphi})$
with the relation $\displaystyle{\left\{Q_{\rm F}, Q_{\rm F}^{\dagger}\right\} = N_{\rm D}}$.
In the case that $\phi$ is invariant under $OSp(2|2)$ transformation,
i.e., $\phi$ is $Q_{\rm F}$ and $Q_{\rm F}^{\dagger}$-singlet 
satisfying $\delta_{\rm F} \phi = 0$ and $\delta_{\rm F}^{\dagger} \phi = 0$,
the full system described by $\mathcal{L}_{\rm T}$ has $OSp(2|2)$ invariance.
Note that $Q_{\rm F}$ and $Q_{\rm F}^{\dagger}$-singlets are allowed,
because $N_{\rm D}$ is irrelevant to spacetime symmetries,
different from the case of spacetime SUSY.

To formulate our model in a consistent manner,
we use a feature
that {\it a conserved charge can, in general, be set to be zero as a subsidiary condition}.
We impose the following subsidiary conditions on states to select physical states
$|{\rm phys}\rangle$ can be selected,\footnote{
The conditions (\ref{Phys}) are interpreted as counterparts of the Kugo-Ojima subsidiary condition
in BRST quantization~\cite{K&O}.
It is shown that the system containing both free ordinary fields and their ghost partners
is quantized consistently, though it becomes empty leaving the vacuum state alone~\cite{YK3}.
}
\begin{eqnarray}
Q_{\rm F} |{\rm phys}\rangle = 0~,~~Q_{\rm F}^{\dagger} |{\rm phys}\rangle = 0~,~~
N_{\rm D} |{\rm phys}\rangle = 0~,
\label{Phys}
\end{eqnarray}
and then heavy fields $\varphi$ and $c_{\varphi}$ are expected to be unphysical
and not to give any quantum effects on the light field $\phi$.
This is regarded as a field theoretical version of the Parisi-Sourlas mechanism~\cite{P&S}.
Hence, there is a possibility that 
$\varphi$ and $c_{\varphi}$ are not dangerous for $\phi$,
and vice versa, and the structure of theory is stabilized
owing to the fermionic symmetries with a charmed life. 

If we take $\varphi^{\dagger} \varphi \varphi^{\dagger} \varphi$
in place of $\varphi^{\dagger} \varphi \star \varphi^{\dagger} \varphi$ in (\ref{L-T-again}),
the self-interactions of $Q_{\rm F}$-doublet 
$\lambda_{\varphi} \left(\varphi^{\dagger} \varphi + c_{\varphi}^{\dagger} c_\varphi\right)
\star \left(\varphi^{\dagger} \varphi + c_{\varphi}^{\dagger} c_\varphi\right)$ 
in $\mathcal{L}_{\varphi, c}$ is replaced by
\begin{eqnarray}
\lambda_{\varphi} \left(\varphi^{\dagger} \varphi \varphi^{\dagger} \varphi
+ \varphi^{\dagger} \varphi c_{\varphi}^{\dagger} c_\varphi
+ c_{\varphi}^{\dagger} \varphi \varphi^{\dagger} c_{\varphi}
+ \varphi^{\dagger} \varphi \star c_{\varphi}^{\dagger} c_\varphi
- c_{\varphi}^{\dagger} \varphi \star \varphi^{\dagger} c_{\varphi}
+ c_{\varphi}^{\dagger} c_{\varphi} \star  c_{\varphi}^{\dagger} c_{\varphi}\right)~.
\label{L-varphi-c-ordinary}
\end{eqnarray}
Hereafter, we do not consider self-interactions containing both local and non-local ones 
such as (\ref{L-varphi-c-ordinary}),
because the form of these interactions could not be stable 
against radiative corrections in the framework of effective field theory.

More powerful fermionic symmetries might be needed to construct a realistic model.
We, however, do not specify them,
because we have not known what an ultimate theory is
and what underlying symmetries are.
Hence, in the following, we use particle-ghost symmetries
as an illustrative example.

We consider theories with 
fermionic symmetries  (whose generators are denoted by
$Q_{\rm F}$ and $Q_{\rm F}^{\dagger}$) 
that relate particles to their ghost partners
and the bosonic symmetry relating the charge $N_{\rm D}$ for the doublets
(pairs of particles and their ghost partners)
with the relation $\displaystyle{\left\{Q_{\rm F}, Q_{\rm F}^{\dagger}\right\} = N_{\rm D}}$,
and assume that those symmetries are not broken down at the quantum level,
and ghost fields are unphysical and harmless.
Then we arrive at the conjecture that
{\it the gauge hierarchy problem does not occur
and the physical low-energy theory can be described by SM + $\alpha$, 
if a full effective theory has fermionic symmetries with an eternal life,
the SM particles and some extra light fields are $Q_{\rm F}$-singlets
and others including heavy fields are $Q_{\rm F}$-doublets}.
A theory beyond and behind the SM
is expected to be expressed as
\begin{eqnarray}
\mathcal{L}_{\rm BSM} = \mathcal{L}_{\rm light} 
+ \mathcal{L}_{\rm heavy} + \mathcal{L}_{\rm mix}~,~~
\mathcal{L}_{\rm light} 
=  \mathcal{L}_{{\rm SM}+\alpha} 
+ {\bm \delta}_{\rm F} {\bm \delta}_{\rm F}^{\dagger} \Delta\mathcal{L}~,
\label{L-BSM}
\end{eqnarray}
where $\mathcal{L}_{\rm light}$, $\mathcal{L}_{\rm heavy}$
and $\mathcal{L}_{\rm mix}$ stand for the Lagrangian densities of 
the parts $X_{\rm light}$, $X_{\rm heavy}$
and $X_{\rm mix}$, respectively.
The $\mathcal{L}_{{\rm SM}+\alpha}$ represents the Lagrangian densities of SM + $\alpha$,
and ${\bm \delta}_{\rm F} {\bm \delta}_{\rm F}^{\dagger} \Delta\mathcal{L}$ 
contains light $Q_{\rm F}$-doublet fields.
Both $\mathcal{L}_{\rm heavy}$ and $\mathcal{L}_{\rm mix}$ are 
also written in the ${\bm \delta}_{\rm F}$-exact form, 
for instance,
\begin{eqnarray}
&~& \mathcal{L}_{\rm heavy} =
{\bm \delta}_{\rm F}\left[\sum_k \left({c}_{{\rm L}k}^{\dagger} i \overline{\sigma}^{\mu} 
D_{\mu} \psi_{{\rm L}k}
+ {c}_{{\rm R}k}^{\dagger} i \sigma^{\mu} D_{\mu} \psi_{{\rm R}k} 
-M_k {c}_{{\rm L}k}^{\dagger} \psi_{{\rm R}k} - 
M_k {c}_{{\rm R}k}^{\dagger} \psi_{{\rm L}k}\right) \right.
\nonumber \\
&~& ~~~~~~~~~~~~~~~~ 
\left. + \sum_l \left(\left(D^{\mu} {c}_{l}\right)^{\dagger} \left(D_{\mu} \varphi_{l}\right)
- M_l^2  c_{l}^{\dagger} \varphi_{l}\right)
+ \cdots\right]
\nonumber \\
&~& ~~~~~~~~~~~~~ 
= {\bm \delta}_{\rm F} {\bm \delta}_{\rm F}^{\dagger}
\left[\sum_k \left({\psi}_{{\rm L}k}^{\dagger} i \overline{\sigma}^{\mu} 
D_{\mu} \psi_{{\rm L}k}
+ {\psi}_{{\rm R}k}^{\dagger} i \sigma^{\mu} D_{\mu} \psi_{{\rm R}k} 
- M_k {\psi}_{{\rm L}k}^{\dagger} \psi_{{\rm R}k} 
- M_k {\psi}_{{\rm R}k}^{\dagger} \psi_{{\rm L}k}\right) \right.
\nonumber \\
&~& ~~~~~~~~~~~~~~~~ 
\left. + \sum_l \left(\left(D^{\mu} \varphi_{l}\right)^{\dagger} \left(D_{\mu} \varphi_{l}\right)
- M_l^2 \varphi_{l}^{\dagger} \varphi_{l}\right)+ \cdots\right]~,
\label{L-High2}\\
&~& \mathcal{L}_{\rm mix}
= {\bm \delta}_{\rm F}\left[-\sum_l \lambda_{l} H^{\dagger} H {c}_{l}^{\dagger}\varphi_{l} 
+ \cdots\right]
= {\bm \delta}_{\rm F} {\bm \delta}_{\rm F}^{\dagger}
\left[-\sum_l \lambda_{l} H^{\dagger} H \varphi_{l}^{\dagger}\varphi_{l} + \cdots\right]~,
\label{L-int2}
\end{eqnarray}
where $(\psi_{{\rm L}k}, c_{{\rm L}k})$ and $(\psi_{{\rm R}k}, c_{{\rm R}k})$
are heavy Weyl spinor $Q_{\rm F}$-doublets,
and $(\varphi_l, c_{l})$ are complex scalar $Q_{\rm F}$-doublets,
and $H$ is the Higgs boson in the SM that is a $Q_{\rm F}$-singlet.
Note that both $\mathcal{L}_{\rm heavy}$ and $\mathcal{L}_{\rm mix}$ are also 
expressed in the ${\bm \delta}^{\dagger}_{\rm F}$-exact form.

\subsection{Grand unification and fermionic symmetry}

First, let us start with a conjecture relating an ultimate theory.
The ultimate theory must explain the birth of our physical world as follows.
{\it Our world comes into existence from $\lq\lq$nothing''.
Here, nothing means not an empty but a world 
whose constituents are unphysical objects
and/or fundamental objects including only gauge bosons (and their superpartners, i.e., gauginos),
that form multiplets of a large gauge symmetry.
$\lq\lq$Beings'' including matter fields are generated at $M_{\rm U}$, 
after reducing the large symmetry
into a smaller one, by some mechanism.
The constituents after the reduction 
are massless particles and massive unphysical ones.
The massless particles contain GUT multiplets and incomplete ones.
Parts of the GUT multiplets
become unphysical in collaboration with ghost partners belonging to incomplete ones.
After all, the SM particles and extra particles survive as physical ones or $\lq\lq$beings'', 
in a lower-energy world.}\footnote{
Based on this conjecture, a toy model has been proposed
that physical modes are released from unobservable fields~\cite{YK4}.
}
Note that, in our scenario, extra components of GUT multiplets can remain massless and decouple to the SM particles
because they become unphysical,
which is different from the ordinary case that they decouple to the SM particles 
because they become heavy on the breakdown of GUT symmetry.

Next, we discuss the verifiability and predictions of our conjecture.
Although unphysical particles do not give any dynamical effects on the physical sector, 
there are at least two predictions, which can be indirect proofs of unphysical sector.
First one is that physical quantities calculated in the SM + $\alpha$ should precisely 
match with the experimental values at the terascale, up to any gravitational effects, 
because radiative corrections from unphysical particles are canceled out.
Second one is that parameters in the SM + $\alpha$ should satisfy specific relations 
at $M_{\rm U}$, reflecting on a large symmetry realized in the ultimate theory.

In the following subsections, we explain that the second prediction can be realized, 
in the appearance of new particles ($Q_{\rm F}$-singlets) around the terascale
and light $Q_{\rm F}$-doublets,
under a situation with following features.\\
(i) An ultimate theory has a large gauge symmetry potentially.
Gauge bosons originate from an object
such as $D$-brane in string theory.\\
(ii) Other particles including matter fields appear
with changing the structure of space-time and/or object at a high-energy scale $M_{\rm U}$.
All massive fields form $Q_{\rm F}$-doublets and become unphysical.
Massless fields consist of ordinary fields and ghost fields.
Most ordinary fields including the gauge bosons form multiplets of a gauge group $G_{\rm o}$,
other ordinary fields form multiplets of a smaller gauge group $G'_{\rm o}$
and ghost fields form multiplets of a gauge group $G_{\rm g}$.
The gauge symmetry of the system could increase 
from  $G'_{\rm o}$ or $G_{\rm g}$ to $G_{\rm o}$, 
if other ordinary fields and massless ghost fields were removed, 
i.e., $G_{\rm o} \supset G'_{\rm o}, G_{\rm g}$.\\
(iii) The system survives in a consistent manner, thanks to fermionic symmetries.
The fermionic symmetries are unbroken at the quantum level,
and all ghost fields are unphysical and harmless.

\subsubsection{Symmetry reduction with ghost administration}
\label{Symmetry reduction with ghost administration}

Let us demonstrate that the gauge symmetry is reduced 
in the appearance of incomplete multiplets at $M_{\rm U}$,
using toy models with the $SU(2)$ Yang-Mills field.

We assume that the ultimate theory possesses many solutions
corresponding multiverse
such as the string landscape 
and some solutions contain ghost fields.
Their low-energy effective field theories are constructed
from the massless spectra.
In the following, we write down the Lagrangian densities
with particle-ghost symmetries if ghost fields exist,
in several cases
for a given set of massless particles.

First we consider an ordinary case that there are no ghost fields.\\
(A) Case with $Q_{\rm F}$ singlets matter fields\\
Let the set $(A_{\mu}^a, \phi, \psi_{\rm L}, \psi_{\rm R})$ 
be given as the massless ones.
Here, $A_{\mu}^a$ are $SU(2)$ gauge bosons $(a=1,2,3)$, 
$\phi =(\phi^1, \phi^2)^T$ is a scalar field of $SU(2)$ doublet
(the superscript $T$ represents the operation of transposition),
$\psi_{\rm L} =(\psi_{\rm L}^1, \psi_{\rm L}^2)^T$ 
and $\psi_{\rm R}=(\psi_{\rm R}^1, \psi_{\rm R}^2)^T$ 
are left-handed and right-handed chiral fermions of $SU(2)$ doublets, respectively.
From the $SU(2)$ gauge invariance, the Lagrangian density is given by
\begin{eqnarray}
&~& \mathcal{L}_{SU(2)}^{\rm (A)} = - \frac{1}{4} F_{\mu\nu}^{a} F^{a\mu\nu}
+ \mathcal{L}_{\rm m}~,
\label{L-SU2-A}\\
&~& \mathcal{L}_{\rm m} 
= (D_{\mu} \phi)^{\dagger} (D^{\mu} \phi) 
+ {\psi}_{\rm L}^{\dagger} i \overline{\sigma}^{\mu} D_{\mu} \psi_{\rm L} 
+ {\psi}_{\rm R}^{\dagger} i \sigma^{\mu} D_{\mu} \psi_{\rm R}~,
\label{L-M}
\end{eqnarray}
where $F_{\mu\nu}^{a} = \partial_{\mu} A_{\nu}^a - \partial_{\nu} A_{\mu}^a 
- g \varepsilon^{abc} A_{\mu}^{b} A_{\nu}^{c}$,
$D_{\mu} = \partial_{\mu} + igA_{\mu}^a \tau^a/2$,
and $g$ is a gauge coupling constant, $\tau^a$ are Pauli matrices.
For simplicity, we omit interactions other than gauge interactions.
The system is described by an ordinary $SU(2)$ Yang-Mills theory
with a complex scalar field and two Weyl spinors (a Dirac spinor).

Next we consider the extremal case that all matter fields company with their ghost partners.\\
(B) Case with $Q_{\rm F}$ doublets matter fields\\
Let the set $(A_{\mu}^a; \phi, c_{\phi}; \psi_{\rm L}, c_{\rm L}; \psi_{\rm R}, c_{\rm R})$ 
be given as the massless ones.
Here, $c_{\phi}$ is the ghost partner of $\phi$,
and $c_{\rm L}$ and $c_{\rm R}$ are the ghost partners of
$\psi_{\rm L}$ and $\psi_{\rm R}$, respectively.
To formulate a theory consistently, we require the $SU(2)$ gauge invariance
and the invariance 
under the fermionic transformations,
\begin{eqnarray}
\hspace{-1.4cm} &~& {\bm \delta}_{\rm F} \phi = - c_{\phi}~,~~
{\bm \delta}_{\rm F} \phi^{\dagger} = 0~,~~ 
{\bm \delta}_{\rm F} c_{\phi} = 0~,~~ 
{\bm \delta}_{\rm F} c_{\phi}^{\dagger} = \phi^{\dagger}~,~~
{\bm \delta}_{\rm F} \psi_{\rm L}= -c_{\rm L}~,~~
{\bm \delta}_{\rm F} \psi_{\rm L}^{\dagger} = 0~,~~ 
\nonumber \\
\hspace{-1.4cm} &~&
{\bm \delta}_{\rm F} c_{\rm L} = 0~,~~ 
{\bm \delta}_{\rm F} c_{\rm L}^{\dagger} = -\psi_{\rm L}^{\dagger}~,~~
{\bm \delta}_{\rm F} \psi_{\rm R} = -c_{\rm R}~,~~
{\bm \delta}_{\rm F} \psi_{\rm R}^{\dagger} = 0~,~~ 
{\bm \delta}_{\rm F} c_{\rm R} = 0~,~~
{\bm \delta}_{\rm F} c_{\rm R}^{\dagger} = -\psi_{\rm R}^{\dagger}~,~~
{\bm \delta}_{\rm F} A_{\mu}^a = 0
\label{delta-F-B}
\end{eqnarray}
and 
\begin{eqnarray}
\hspace{-1.4cm} &~& {\bm \delta}_{\rm F}^{\dagger} \phi = 0~,~~
{\bm \delta}_{\rm F}^{\dagger} \phi^{\dagger} =c_{\phi}^{\dagger}~,~~ 
{\bm \delta}_{\rm F}^{\dagger} c_{\phi} = \phi~,~~ 
{\bm \delta}_{\rm F}^{\dagger} c_{\phi}^{\dagger} = 0~,~~
{\bm \delta}_{\rm F}^{\dagger} \psi_{\rm L}= 0~,~~
{\bm \delta}_{\rm F}^{\dagger} \psi_{\rm L}^{\dagger} = -c_{\rm L}^{\dagger}~,~~ 
\nonumber \\
\hspace{-1.4cm} &~&
{\bm \delta}_{\rm F}^{\dagger} c_{\rm L} = \psi_{\rm L}~,~~ 
{\bm \delta}_{\rm F}^{\dagger} c_{\rm L}^{\dagger} = 0~,~~
{\bm \delta}_{\rm F}^{\dagger} \psi_{\rm R} = 0~,~~
{\bm \delta}_{\rm F}^{\dagger} \psi_{\rm R}^{\dagger} = -c_{\rm R}^{\dagger}~,~~ 
{\bm \delta}_{\rm F}^{\dagger} c_{\rm R} = \psi_{\rm R}~,~~
{\bm \delta}_{\rm F}^{\dagger} c_{\rm R}^{\dagger} = 0~,~~
{\bm \delta}_{\rm F}^{\dagger} A_{\mu}^a = 0~.
\label{delta-F-dagger-B}
\end{eqnarray}
Then, we obtain the Lagrangian density
\begin{eqnarray}
&~& \mathcal{L}_{SU(2)}^{\rm (B)} = - \frac{1}{4} F_{\mu\nu}^{a} F^{a\mu\nu}
+ \mathcal{L}_{\rm m} + \mathcal{L}_{\rm gh}^{\rm (B)}
= - \frac{1}{4} F_{\mu\nu}^{a} F^{a\mu\nu} 
+ {\bm \delta}_{\rm F} {\bm \delta}_{\rm F}^{\dagger} \mathcal{L}_{\rm m}~,
\label{L-SU2-B}\\
&~&  \mathcal{L}_{\rm gh}^{\rm (B)}
= (D_{\mu} c_{\phi})^{\dagger} (D^{\mu} c_{\phi}) 
+ {c}_{\rm L}^{\dagger} i \overline{\sigma}^{\mu} D_{\mu} c_{\rm L} 
+ {c}_{\rm R}^{\dagger} i \overline{\sigma}^{\mu} D_{\mu} c_{\rm R} ~.
\label{L-gh-B}
\end{eqnarray}
For simplicity, we omit interactions other than gauge interactions.
The system is essentially identical to that described by the pure $SU(2)$ Yang-Mills theory,
because $Q_{\rm F}$ doublets are unphysical under the subsidiary conditions (\ref{Phys}).

Here, we give a comment on a SUSY extension of the system.
Let the set $(A_{\mu}^a; \lambda^a, c^a)$ be given as the massless ones.
Here, $\lambda^a$ are $SU(2)$ gauginos and $c^a$ are their ghost partners.
We obtain the Lagrangian density
\begin{eqnarray}
&~& \mathcal{L}_{SU(2)}^{\rm (B')} = - \frac{1}{4} F_{\mu\nu}^{a} F^{a\mu\nu}
+ \frac{1}{2} \overline{\lambda}^{a}  i \gamma^{\mu} (D_{\mu} \lambda)^a
+ \frac{1}{2} \overline{c}^{a}  i \gamma^{\mu} (D_{\mu} c)^a
\nonumber \\
&~& ~~~~~~~~~~~~~~ = - \frac{1}{4} F_{\mu\nu}^{a} F^{a\mu\nu} 
+ {\bm \delta}_{\rm F} {\bm \delta}_{\rm F}^{\dagger} 
\left(\frac{1}{2} \overline{\lambda}^{a}  i \gamma^{\mu} (D_{\mu} \lambda)^a\right)~.
\label{L-SU2-B'}
\end{eqnarray}
This system is also identical to that described by the pure $SU(2)$ Yang-Mills theory,
under the subsidiary conditions (\ref{Phys}).

Finally, we consider an exotic case such that incomplete ghost fields exist.\\
(C) Case with incomplete $Q_{\rm F}$ singlets matter fields\\
Let us obtain the set of particles $A_{\mu}^a$, $\phi$, 
$\psi_{\rm L}$, $\psi_{\rm R}$ 
and the ghost fields which do not form $SU(2)$ multiplets
such as $C_{\mu}^{+}$, $C_{\mu}^{-}$, $c_{\phi}^1$, $c_{\rm L}^1$ and $c_{\rm R}^1$,
as the massless ones.
The gauge quantum numbers of ghost fields are same as those of
$A_{\mu}^{+}=(A_{\mu}^1 - i A_{\mu}^2)/\sqrt{2}$, 
$A_{\mu}^{-}=(A_{\mu}^1 + i A_{\mu}^2)/\sqrt{2}$, ${\phi}^1$, 
$\psi_{\rm L}^1$ and $\psi_{\rm R}^1$, respectively,
but they obey statistics different from ordinary counterparts.
To formulate a theory, we require the $U(1)$ gauge invariance
and the invariance 
under the fermionic transformations,
\begin{eqnarray}
&~& {\bm \delta}_{\rm F} \phi^1 = - c_{\phi}^1~,~~
{\bm \delta}_{\rm F} \phi^{1\dagger} = 0~,~~ 
{\bm \delta}_{\rm F} c_{\phi}^1 = 0~,~~ 
{\bm \delta}_{\rm F} c_{\phi}^{1\dagger} = \phi^{1\dagger}~,~~
{\bm \delta}_{\rm F} \psi_{\rm L}^1 
= -c_{\rm L}^1~,~~{\bm \delta}_{\rm F} \psi_{\rm L}^{1\dagger} = 0~,~~
\nonumber \\ 
&~& {\bm \delta}_{\rm F} c_{\rm L}^1 = 0~,~~ 
{\bm \delta}_{\rm F} c_{\rm L}^{1\dagger} = -\psi_{\rm L}^{1\dagger}~,~~
{\bm \delta}_{\rm F} \psi_{\rm R}^1 
= -c_{\rm R}^1~,~~{\bm \delta}_{\rm F} \psi_{\rm R}^{1\dagger} = 0~,~~ 
{\bm \delta}_{\rm F} c_{\rm R}^1 = 0~,~~ 
{\bm \delta}_{\rm F} c_{\rm R}^{1\dagger} = -\psi_{\rm R}^{1\dagger}~,~~
\nonumber \\
&~& {\bm \delta}_{\rm F} A_{\mu}^+ 
= - C_{\mu}^+~,~~{\bm \delta}_{\rm F} A_{\mu}^{-} = 0~,~~ 
{\bm \delta}_{\rm F} C_{\mu}^+ = 0~,~~ {\bm \delta}_{\rm F} C_{\mu}^{-} = A_{\mu}^{-}~,~~
{\bm \delta}_{\rm F} \phi^2 = 0~,~~ {\bm \delta}_{\rm F} \phi^{2\dagger} = 0~,~~
\nonumber \\
&~& {\bm \delta}_{\rm F} \psi_{\rm L}^2 = 0~,~~ {\bm \delta}_{\rm F} \psi_{\rm L}^{2\dagger}=0~,~~
{\bm \delta}_{\rm F} \psi_{\rm R}^2 = 0~,~~ 
{\bm \delta}_{\rm F} \psi_{\rm R}^{2\dagger} = 0~,~~
{\bm \delta}_{\rm F} A_{\mu}^{3} = 0
\label{delta-F-C}
\end{eqnarray}
and
\begin{eqnarray}
&~& {\bm \delta}_{\rm F}^{\dagger} \phi^1 = 0~,~~
{\bm \delta}_{\rm F}^{\dagger} \phi^{1\dagger} = c_{\phi}^{1\dagger}~,~~ 
{\bm \delta}_{\rm F}^{\dagger} c_{\phi}^1 = \phi^1~,~~ 
{\bm \delta}_{\rm F}^{\dagger} c_{\phi}^{1\dagger} = 0~,~~
{\bm \delta}_{\rm F}^{\dagger} \psi_{\rm L}^1 = 0~,~~
{\bm \delta}_{\rm F}^{\dagger} \psi_{\rm L}^{1\dagger} =  -c_{\rm L}^{1\dagger}~,~~
\nonumber \\ 
&~& {\bm \delta}_{\rm F}^{\dagger} c_{\rm L}^1 = \psi_{\rm L}^{1}~,~~ 
{\bm \delta}_{\rm F}^{\dagger} c_{\rm L}^{1\dagger} = 0~,~~
{\bm \delta}_{\rm F}^{\dagger} \psi_{\rm R}^1 = 0~,~~
{\bm \delta}_{\rm F}^{\dagger} \psi_{\rm R}^{1\dagger} = -c_{\rm R}^{1\dagger}~,~~ 
{\bm \delta}_{\rm F}^{\dagger} c_{\rm R}^1 = \psi_{\rm R}^{1}~,~~ 
{\bm \delta}_{\rm F}^{\dagger} c_{\rm R}^{1\dagger} = 0~,~~
\nonumber \\
&~& {\bm \delta}_{\rm F}^{\dagger} A_{\mu}^+ = 0~,~~
{\bm \delta}_{\rm F}^{\dagger} A_{\mu}^{-} = C_{\mu}^{-}~,~~ 
{\bm \delta}_{\rm F}^{\dagger} C_{\mu}^+ = A_{\mu}^{+}~,~~ 
{\bm \delta}_{\rm F}^{\dagger} C_{\mu}^{-} = 0~,~~
{\bm \delta}_{\rm F}^{\dagger} \phi^2 = 0~,~~ 
{\bm \delta}_{\rm F}^{\dagger} \phi^{2\dagger} = 0~,~~
\nonumber \\
&~& {\bm \delta}_{\rm F}^{\dagger} \psi_{\rm L}^2 = 0~,~~ 
{\bm \delta}_{\rm F}^{\dagger} \psi_{\rm L}^{2\dagger}=0~,~~
{\bm \delta}_{\rm F}^{\dagger} \psi_{\rm R}^2 = 0~,~~ 
{\bm \delta}_{\rm F}^{\dagger} \psi_{\rm R}^{2\dagger} = 0~,~~
{\bm \delta}_{\rm F}^{\dagger} A_{\mu}^{3} = 0~.
\label{delta-F-dagger-C}
\end{eqnarray}
Then, we obtain the Lagrangian density,
\begin{eqnarray}
&~& \mathcal{L}_{SU(2)}^{\rm (C)} = - \frac{1}{4} [F_{\mu\nu}^{a} F^{a\mu\nu}]_{\star}
+ \mathcal{L}_{\rm m} + \mathcal{L}_{\rm gh}^{\rm (C)} + \mathcal{L}_{\rm int}^{\rm (C)}~,
\label{L-SU2-C}
\end{eqnarray}
where 
$[F_{\mu\nu}^{a} F^{a\mu\nu}]_{\star}$ is the gauge kinetic term
that $(g^2/4) (A_{\nu}^- A_{\mu}^+ - A_{\mu}^- A_{\nu}^+) 
(A^{-\nu}A^{+\mu} - A^{-\mu}A^{+\nu})$
is replaced by the non-local one
$(g^2/4) (A_{\nu}^- A_{\mu}^+ - A_{\mu}^- A_{\nu}^+) \star
(A^{-\nu}A^{+\mu} - A^{-\mu}A^{+\nu})$ in $F_{\mu\nu}^{a} F^{a\mu\nu}$,
and $\mathcal{L}_{\rm gh}^{\rm (C)}$ and $\mathcal{L}_{\rm int}^{\rm (C)}$ are given by
\begin{eqnarray}
&~& \mathcal{L}_{\rm gh}^{\rm (C)} = 
- (D'_{\mu} C_{\nu}^-)(D'^{\mu} C^{+\nu}) + (D'_{\mu} C_{\nu}^-)(D'^{\nu} C^{+\mu})
+ (D'_{\mu} c_{\phi}^1)^{\dagger} (D'^{\mu} c_{\phi}^1)
\nonumber \\
&~& ~~~~~~~~~~~~~~
+ c_{\rm L}^{1\dagger} i \overline{\sigma}^{\mu} D'_{\mu} c_{\rm L}^1
+ c_{\rm R}^{1\dagger} i \sigma^{\mu} D'_{\mu} c_{\rm R}^1 ~,
\label{L-gh-C}\\
&~& \mathcal{L}_{\rm int}^{\rm (C)} 
= - \frac{ig}{2}\left(\partial_{\mu} A_{\nu}^3 - \partial_{\nu} A_{\mu}^3\right)
\left(C^{-\nu}C^{+\mu} - C^{-\mu}C^{+\nu}\right)
\nonumber \\
&~& ~~~~~~~~~~~~~~
+ \frac{g^2}{2}\left(A_{\nu}^- A_{\mu}^+ - A_{\mu}^- A_{\nu}^+\right) \star
\left(C^{-\nu}C^{+\mu} - C^{-\mu}C^{+\nu}\right)
\nonumber \\
&~& ~~~~~~~~~~~~~~
 + \frac{g^2}{4}\left(C_{\nu}^- C_{\mu}^+ - C_{\mu}^- C_{\nu}^+\right) \star
\left(C^{-\nu}C^{+\mu} - C^{-\mu}C^{+\nu}\right)
\nonumber \\
&~& ~~~~~~~~~~~~~~
+ \frac{g^2}{2} \left(-\phi^{1\dagger}C_{\mu}^+ C^{-\mu}\phi^1 
+ \phi^{2\dagger} C_{\mu}^- C^{+\mu}\phi^2 \right)
\nonumber \\
&~& ~~~~~~~~~~~~~~ 
+ \frac{g^2}{2} c_{\phi}^{1\dagger}\left(A_{\mu}^+ A^{-\mu} - C_{\mu}^+ C^{-\mu}\right)c_{\phi}^1
+ \frac{ig}{\sqrt{2}} (D'_{\mu} c_{\phi}^1)^{\dagger} C^{+\mu} {\phi}^2 
- \frac{ig}{\sqrt{2}} {\phi}^{2\dagger}C_{\mu}^- (D'^{\mu} c_{\phi}^1)
\nonumber\\
&~& ~~~~~~~~~~~~~~ 
+ \frac{ig}{\sqrt{2}} (D'_{\mu} {\phi}^2)^{\dagger} C^{-\mu} c_{\phi}^1 
- \frac{ig}{\sqrt{2}} c_{\phi}^{1\dagger}C_{\mu}^+ (D'^{\mu} {\phi}^2)
+ \frac{g}{\sqrt{2}} c_{\rm L}^{1\dagger} \overline{\sigma}^{\mu} C_{\mu}^+ \psi_{\rm L}^2
\nonumber \\
&~& ~~~~~~~~~~~~~~ 
+ \frac{g}{\sqrt{2}} \psi_{\rm L}^{2\dagger} \overline{\sigma}^{\mu} C_{\mu}^- c_{\rm L}^1
+ \frac{g}{\sqrt{2}} c_{\rm R}^{1\dagger} \overline{\sigma}^{\mu} C_{\mu}^+ \psi_{\rm R}^2
+ \frac{g}{\sqrt{2}} \psi_{\rm R}^{2\dagger} \overline{\sigma}^{\mu} C_{\mu}^- c_{\rm R}^1~,
\label{L-int-C}
\end{eqnarray}
where $D'_{\mu} = \partial_{\mu} + i g A_{\mu}^3 T^3$ 
($T^3$ is the third component of $su(2)$ algebra), 
$\mathcal{L}_{\rm gh}^{\rm (C)}$ are kinetic terms of ghost fields
including a minimal coupling with the $U(1)$ gauge boson $A_{\mu}^3$,
and $\mathcal{L}_{\rm int}$ contains interactions between ordinary matters and ghosts.

The total Lagrangian density is rewritten as
\begin{eqnarray}
\mathcal{L}_{SU(2)}^{\rm (C)} = - \frac{1}{4} [F_{\mu\nu}^{a} F^{a\mu\nu}]_{\star}
+ \mathcal{L}_{\rm m} + \mathcal{L}_{\rm gh}^{\rm (C)} + \mathcal{L}_{\rm int}^{\rm (C)}
= \mathcal{L}_{U(1)} 
+ {\bm \delta}_{\rm F} {\bm \delta}_{\rm F}^{\dagger} \Delta\mathcal{L}^{\rm (C)}~,
\label{L-T2}
\end{eqnarray}
where $\mathcal{L}_{U(1)}$ and $\Delta\mathcal{L}^{\rm (C)}$ are given by,
\begin{eqnarray}
&~& \mathcal{L}_{U(1)} = - \frac{1}{4} (\partial_{\mu} A_{\nu}^3 - \partial_{\nu} A_{\mu}^3)
(\partial^{\mu} A^{3\nu} - \partial^{\nu} A^{3\mu})
+ (D'_{\mu} \phi^2)^{\dagger} (D'^{\mu} \phi^2)
\nonumber \\
&~& ~~~~~~~~~~~~~~~~
+ {\psi}_{\rm L}^{2\dagger} i \overline{\sigma}^{\mu} D'_{\mu} \psi_{\rm L}^2
+ {\psi}_{\rm R}^{2\dagger} i \sigma^{\mu} D'_{\mu} \psi_{\rm R}^2~,
\label{L-U1}\\
&~& \Delta\mathcal{L}^{\rm (C)} = 
- (D'_{\mu} A_{\nu}^-)(D'^{\mu} A^{+\nu}) + (D'_{\mu} A_{\nu}^-)(D'^{\nu} A^{+\mu})
\nonumber \\
&~& ~~~~~~~~~
- \frac{ig}{2}\left(\partial_{\mu} A_{\nu}^3 - \partial_{\nu} A_{\mu}^3\right)
\left(A^{-\nu}A^{+\mu} - A^{-\mu}A^{+\nu}\right)
\nonumber \\
&~& ~~~~~~~~~ 
+ \frac{g^2}{8}\left(A_{\nu}^- A_{\mu}^+ - A_{\mu}^- A_{\nu}^+\right) \star
\left(A^{-\nu}A^{+\mu} - A^{-\mu}A^{+\nu}\right)
\nonumber \\
&~& ~~~~~~~~~ 
- \frac{g^2}{8}\left(C_{\nu}^- C_{\mu}^+ - C_{\mu}^- C_{\nu}^+\right) \star
\left(C^{-\nu}C^{+\mu} - C^{-\mu}C^{+\nu}\right)
+ (D'_{\mu} {\phi}^1)^{\dagger} (D'^{\mu} {\phi}^1)
\nonumber \\
&~& ~~~~~~~~~ 
+ \frac{g^2}{4} \left(\phi^{1\dagger} A_{\mu}^+ A^{-\mu} \phi^1 
+ c_{\phi}^{1\dagger} C_{\mu}^+ C^{-\mu} c_{\phi}^1\right)
+ \frac{g^2}{2} \phi^{2\dagger} A^{-\mu} A_{\mu}^+ \phi^2
\nonumber \\
&~& ~~~~~~~~~ 
+ \frac{ig}{\sqrt{2}}(D'_{\mu} {\phi}^1)^{\dagger} A^{+\mu} {\phi}^2 
- \frac{ig}{\sqrt{2}} {\phi}^{2\dagger} A_{\mu}^- (D'^{\mu} \phi^1)
\nonumber \\
&~& ~~~~~~~~~ + \frac{ig}{\sqrt{2}} (D'_{\mu} {\phi}^2)^{\dagger} A^{-\mu} {\phi}^1 
- \frac{ig}{\sqrt{2}} {\phi}^{1\dagger} A_{\mu}^+ (D'^{\mu} {\phi}^2)
\nonumber \\
&~& ~~~~~~~~~ 
+ \psi_{\rm L}^{1\dagger} i \overline{\sigma}^{\mu} D'_{\mu} \psi_{\rm L}^1
- \frac{g}{\sqrt{2}} \psi_{\rm L}^{1\dagger} \overline{\sigma}^{\mu} A_{\mu}^+ \psi_{\rm L}^2
- \frac{g}{\sqrt{2}} \psi_{\rm L}^{2\dagger} \overline{\sigma}^{\mu} A_{\mu}^- \psi_{\rm L}^1
\nonumber \\
&~& ~~~~~~~~~ 
+ \psi_{\rm R}^{1\dagger} i {\sigma}^{\mu} D'_{\mu} \psi_{\rm R}^1
- \frac{g}{\sqrt{2}} \psi_{\rm R}^{1\dagger} {\sigma}^{\mu} A_{\mu}^+ \psi_{\rm R}^2
- \frac{g}{\sqrt{2}} \psi_{\rm R}^{2\dagger} {\sigma}^{\mu} A_{\mu}^- \psi_{\rm R}^1~.
\label{DeltaL}
\end{eqnarray}
The system is essentially identical to the $U(1)$ gauge theory
described by $\mathcal{L}_{U(1)}$ under the subsidiary conditions (\ref{Phys}).

From (\ref{L-T2}), we find that
$SU(2)$ gauge symmetry is hidden in the form that it emerges after removing ghost fields
and replacing the non-local self-interactions among $A_{\mu}^{\pm}$
by the local ones.
The $A_{\mu}^+$ and $A_{\mu}^-$ behave as charged matters 
and change their phase under the $U(1)$ gauge transformation.
The time-components of $A_{\mu}^{\pm}$ generate negative norm states,
but they can be unphysical and harmless with the help of those of $C_{\mu}^{\pm}$.
Hence the theory would not encounter inconsistency, 
so far as the $U(1)$ gauge invariance and particle-ghost symmetries are respected.
To treat the non-local interactions
and formulate the system consistently,
the framework beyond the effective field theory might be necessary.

Furthermore, (\ref{L-T2}) can be regarded as a matching condition
between a system with $SU(2)$ gauge bosons and that with the reduced $U(1)$
symmetry at a high-energy scale $M_{\rm U}$, where matters and ghosts are administrated.
Hence, we expect that specific relations among physical parameters,
reflecting a larger gauge symmetry, are revived at $M_{\rm U}$, 
and they are tested by analyzing renormalization group flows of parameters 
in a system with a reduced gauge symmetry.

\subsubsection{Grand unification scenario}

We take the following viewpoint and scenario for a physics
beyond and behind the SM.
The gauge coupling constants precisely measured 
at the Large Electron-Positron collider (LEP)~\cite{LEP}
suggest that the SM gauge interactions are unified at $M_{\rm U}$ in SM + $\alpha$.
An ultimate theory has a grand unified gauge symmetry potentially,
and contains both massless and massive states.
All massive states form doublets of $Q_{\rm F}$, and they become unphysical.
Massless states consist of three types of constituents,
ordinary fields (collectively denoted by $\Phi_{\rm U}$) including the gauge bosons 
which belong to multiplets of a unified gauge group $G_{\rm U}$,
ordinary fields (collectively denoted by $\Phi'_{\rm o}$) 
which belong to those of a smaller gauge symmetry $G'_{\rm o}$,
and ghost fields (collectively denoted by $\Phi_{\rm g}$) 
which belong to those of a gauge symmetry $G_{\rm g}$.
The physics of $\Phi_{\rm U}$ is effectively described by a GUT.
If $G'_{\rm o}$ and/or $G_{\rm g}$ is the gauge group of SM + $\alpha$,
the GUT symmetry is broken down into the SM + $\alpha$ one at $M_{\rm U}$, 
in the presence of $\Phi'_{\rm o}$ and $\Phi_{\rm g}$.
Then the theory turns out to be SM + $\alpha$ with specific relations 
among parameters reflecting on the unified symmetry, at $M_{\rm U}$.
Or specific initial conditions are imposed on parameters of SM + $\alpha$, at $M_{\rm U}$.
Note that there are no contributions such as threshold corrections due to heavy particles,
in case that they are unphysical and do not give any quantum effects.

With the help of the toy model (C)
in Sect. \ref{Symmetry reduction with ghost administration},
our scenario is summarized as\footnote{
The basic idea of our scenario is same as those in Refs.~\cite{YK1,YK2}.
}
\begin{eqnarray}
\left. \mathcal{L}_{\rm light} = \mathcal{L}_{{\rm GUT}\star} + \mathcal{L}'_{\rm o} 
+ \mathcal{L}_{\rm gh} + \mathcal{L}_{\rm int}
= \mathcal{L}_{{\rm SM}+\alpha} 
+ {\bm \delta}_{\rm F}  {\bm \delta}_{\rm F}^{\dagger} \Delta\mathcal{L}~~\right|_{M_{\rm U}}~,
\label{L-lihgt}
\end{eqnarray}
where $\mathcal{L}_{{\rm GUT}\star}$ is the Lagrangian density 
describing the GUT concerning $\Phi_{\rm U}$,
$\mathcal{L}'_{\rm o}$ and $\mathcal{L}_{\rm gh}$ contain 
kinetic terms of $\Phi'_{\rm o}$ and $\Phi_{\rm g}$
including minimal couplings with the gauge bosons in SM + $\alpha$,
and $\mathcal{L}_{\rm int}$ contains interactions between ordinary particles and ghosts.
We present a prototype model describing the grand unification, in the Appendix B. 

The theory has following excellent features.
\begin{itemize}
\item The Lagrangian density in SM + $\alpha$ is obtained with the following 
conditions for gauge coupling constants,
\begin{eqnarray}
\left. g_3 = g_2 = g_1 = g_{\rm U}~~\right|_{M_{\rm U}}~,~~ g_1 = \sqrt{\frac{5}{3}} g'~,
\label{SM-ICs}
\end{eqnarray}
where $g_3$, $g_2$ and $g'$ are the gauge coupling constants 
for $SU(3)_{\rm C}$, $SU(2)_{\rm L}$ and $U(1)_{\rm Y}$, respectively,
and $g_{\rm U}$ is the unified gauge coupling constant.
\item The triplet-doublet splitting of Higgs boson is realized,
if extra colored components are unphysical with the advent of their ghost partners.
\item The SM gauge interactions are unified under a large gauge group,
but the proton can be stabilized
if extra colored particles such as $X$ gauge bosons are unphysical,
in the presence of their ghost partners,
and do not give any quantum effects on physical particles.
\end{itemize}

Furthermore, new particles around the terascale in SM + $\alpha$ can provide useful hints
to the physics such as the grand unification and SUSY at $M_{\rm U}$.
For instance, if (part of) new particles form hypermultiplets as remnants of SUSY,
it can be an evidence of (the reduction of) $N=2$ SUSY through the analysis of 
renormalization group evolutions
of parameters~\cite{K}.

\section{Conclusions}
\label{Conclusions}

We have reconsidered the gauge hierarchy problem
from the viewpoint of effective field theories and a high-energy physics,
motivated by the alternative scenario that the SM (modified with massive neutrinos) 
holds up to a high-energy scale such as the Planck scale
and the principle that the hierarchy is stabilized by a symmetry
that should be unbroken in the SM.
We have given a conjecture that 
theories with specific internal fermionic symmetries 
can be free from the gauge hierarchy problem
and become candidates of the physics beyond and/or behind the SM,
and presented a grand unification scenario and its prototype model.

Our consideration is based on the reinterpretation of the gauge hierarchy problem
such that $\lq\lq${\it without spoiling the structure of a high-energy physics 
supported by an excellent concept,
is it possible to construct a low-energy effective theory 
and the interaction with heavy particles?}"

It is also based on following thoughts.
A large symmetry is, in general, broken down to the smaller one, 
if two systems with different size of symmetries interact with each other.
The spacetime SUSY is no exception.
A requirement of large and manifest symmetries causes strict laws of physics, 
and often leads to an unrealistic system.
Diversity of nature might be a result of a partial breakdown or reduction of such symmetries,
keeping its inner beauty.
It would be attractive that the SM particles behave liberally
to the extent permitted by the laws of physics including hidden symmetries.

In this way, spacetime SUSY seems not to be within reach of our direct measurements,
because it is too beautiful and prominent.
However, it does not mean that SUSY is absent in our world, at all.
It is just contrary, and SUSY must exist at an ultimate level,
because it achieves the unification of bosons and fermions, 
and it is deeply connected to the consistency of the theory such as superstring theory.
Then, it can be said that
the existence of fermions is a proof for SUSY.
It is also possible to gain information on SUSY realized at $M_{\rm U}$
from new particles around the terascale~\cite{K}.

A magical ability would be required to keep an inner beauty eternally.
If fermionic symmetries such as the particle-ghost symmetries remain unbroken,
the SM particles could behave liberally as singlets.
In this situation, even if the SM gauge interactions are unified under a large gauge group,
proton can be stable because 
extra colored particles such as $X$ gauge bosons become unphysical.
In other words, proton acquires an eternal life 
as a result of the fact that extra colored particles sell their souls to the ghosts.

Furthermore, a definite discrepancy has not yet been observed between the predictions
in the SM (modified with massive neutrinos) and experimental results,
and this fact might be a proof for the existence of hidden fermionic symmetries
and its related unphysical particles.
The theory can be tested indirectly, using features of symmetries.
In particular,
physical quantities calculated in the SM + $\alpha$ should precisely 
match with the experimental values at the terascale, because radiative corrections 
from unphysical particles are canceled out.
Parameters in the SM + $\alpha$ should satisfy specific relations at $M_U$, 
reflecting on a large symmetry realized in the ultimate theory. 

Our scenario offers a system where the vacuum energy vanishes at $M_{\rm U}$,
because contributions from heavy particles are canceled out
and those from massless particles turn out to be zero
after the quartic divergences are removed.
Our scenario could also have a long life if consistent,
because both spacetime SUSY and internal fermionic symmetry can coexist.
That is, in case that superpartners are discovered,
they can be treated as new particles in SM + $\alpha$.
If some of superpartners were absent,
our fermionic symmetry would have a chance to show up.

In our formulation, there appear
non-local interactions among unphysical particles.
This fact might suggest that fundamental objects are 
not point particles but extended objects,
and a formulation using extended objects should be 
required to describe the interactions consistently.

Even if our particle-ghost symmetries have a weak point,
our conjecture would be survive
that a magical symmetry can play the central role to solve the gauge hierarchy problem,
if the SM particles are singlets and heavy particles belong to non-singlets
of the transformation group,
and the symmetry is unbroken and hidden in the low-energy theory.

It is important to examine whether theories with internal fermionic symmetries
are consistently formulated in a manner to satisfy unitarity and causality.
It is also challenging to study the structure of ultimate theory 
and to derive its low-energy effective theory.
If our world originated from only unphysical objects,
more powerful symmetries would be needed to formulate unphysical theories
including gauge bosons and gravitons,
and the concept of orbifold grand unification~\cite{OGUT1,OGUT2} would be useful
on the reduction of relevant symmetries.

\section*{Acknowledgments}
This work was supported in part by scientific grants from the Ministry of Education, Culture,
Sports, Science and Technology under Grant Nos.~22540272.

\appendix
\section{Non-local interactions and radiative corrections}

We study interactions among unphysical particles,
and radiative corrections on parameters, using a toy model described by the Lagrangian density,
\begin{eqnarray}
&~& \tilde{\mathcal{L}}_{\rm T} = \mathcal{L}_{\phi} + \tilde{\mathcal{L}}_{\varphi, c} + \tilde{\mathcal{L}}_{\rm mix}~,
\label{L-T-tilde}\\
&~& \mathcal{L}_{\phi} = \partial_\mu \phi^{\dagger} \partial^{\mu} \phi 
- m_{\phi}^2 \phi^{\dagger}\phi
- \lambda_{\phi} \left(\phi^{\dagger}\phi\right)^2~,
\label{L-phi2}\\
&~& \tilde{\mathcal{L}}_{\varphi, c} 
= \partial_\mu \varphi^{\dagger} \partial^{\mu} \varphi 
+ \partial_\mu c_{\varphi}^{\dagger} \partial^{\mu} c_\varphi
- M_{\varphi}^{(\varphi)2} \varphi^{\dagger} \varphi 
- M_{\varphi}^{(c)2} c_{\varphi}^{\dagger} c_\varphi
\nonumber \\
&~& ~~~~~~~~~~~~~~
- \lambda_{\varphi}^{(\varphi)} \varphi^{\dagger} \varphi \star \varphi^{\dagger} \varphi
- 2\lambda_{\varphi}^{(\varphi,c)} \varphi^{\dagger} \varphi \star c_{\varphi}^{\dagger} c_\varphi
- \lambda_{\varphi}^{(c)} c_{\varphi}^{\dagger} c_\varphi \star c_{\varphi}^{\dagger} c_\varphi~,
\label{L-varpsi-tilde}\\
&~& \tilde{\mathcal{L}}_{\rm mix}
= -\lambda'^{(\varphi)} \phi^{\dagger} \phi \varphi^{\dagger} \varphi 
- \lambda'^{(c)} \phi^{\dagger} \phi c_{\varphi}^{\dagger} c_\varphi~,
\label{L-mix-varpsi-tilde}
\end{eqnarray}
where $\lambda_{\phi}$, $\lambda_{\varphi}^{(\varphi)}$ and $\lambda_{\varphi}^{(c)}$ 
are the quartic self-coupling constants of $\phi$, $\varphi$ and $c_{\varphi}$, respectively,
and $\lambda_{\varphi}^{(\varphi,c)}$, $\lambda'^{(\varphi)}$ and $\lambda'^{(c)}$ 
are the quartic coupling constants among $\varphi$, $c_{\varphi}$ and $\phi$.

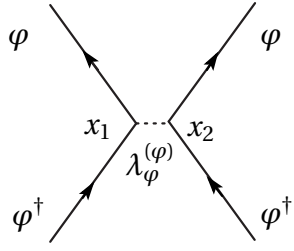
\begin{figure}[hbtp]
\caption[Vertex]{The vertex representing the non-local interaction.}
\label{Fvertex}
\begin{center}
\unitlength 0.1in
\begin{picture}( 13.2000, 12.5000)(  1.8000,-16.4000)
%
\special{pn 13}%
\special{pa 400 400}%
\special{pa 850 1020}%
\special{fp}%
%
\special{pn 13}%
\special{pa 850 1020}%
\special{pa 400 1640}%
\special{fp}%
%
\special{pn 13}%
\special{pa 1470 1630}%
\special{pa 1020 1010}%
\special{fp}%
%
\special{pn 13}%
\special{pa 1020 1010}%
\special{pa 1472 390}%
\special{fp}%
%
\special{pn 13}%
\special{pa 850 1020}%
\special{pa 1010 1020}%
\special{dt 0.045}%
%
\special{pn 13}%
\special{pa 650 740}%
\special{pa 580 650}%
\special{fp}%
\special{sh 1}%
\special{pa 580 650}%
\special{pa 606 716}%
\special{pa 614 692}%
\special{pa 638 690}%
\special{pa 580 650}%
\special{fp}%
%
\special{pn 13}%
\special{pa 570 1410}%
\special{pa 620 1340}%
\special{fp}%
\special{sh 1}%
\special{pa 620 1340}%
\special{pa 566 1384}%
\special{pa 590 1384}%
\special{pa 598 1406}%
\special{pa 620 1340}%
\special{fp}%
%
\special{pn 13}%
\special{pa 1300 1400}%
\special{pa 1230 1310}%
\special{fp}%
\special{sh 1}%
\special{pa 1230 1310}%
\special{pa 1256 1376}%
\special{pa 1264 1352}%
\special{pa 1288 1350}%
\special{pa 1230 1310}%
\special{fp}%
%
\special{pn 13}%
\special{pa 1220 740}%
\special{pa 1270 660}%
\special{fp}%
\special{sh 1}%
\special{pa 1270 660}%
\special{pa 1218 706}%
\special{pa 1242 706}%
\special{pa 1252 728}%
\special{pa 1270 660}%
\special{fp}%
\put(5.7000,-11.0000){\makebox(0,0)[lb]{$x_1$}}%
\put(11.1000,-11.1000){\makebox(0,0)[lb]{$x_2$}}%
\put(2.0000,-15.9000){\makebox(0,0)[lb]{$\varphi^{\dagger}$}}%
\put(15.0000,-15.8000){\makebox(0,0)[lb]{$\varphi^{\dagger}$}}%
\put(1.8000,-6.4000){\makebox(0,0)[lb]{$\varphi$}}%
\put(15.0000,-6.4000){\makebox(0,0)[lb]{$\varphi$}}%
\put(7.9000,-13.5000){\makebox(0,0)[lb]{$\lambda_{\varphi}^{(\varphi)}$}}%
\end{picture}%
\end{center}
\end{figure}

In the form of action integral, the star product ($\star$) represents 
a non-local interaction such that 
$-\lambda_{\varphi}^{(\varphi)} \varphi^{\dagger} \varphi \star \varphi^{\dagger} \varphi$
stands for
\begin{eqnarray}
-\int \lambda_{\varphi}^{(\varphi)} v(x_1, x_2) 
\varphi^{\dagger}(x_1) \varphi(x_1)  \varphi^{\dagger}(x_2) \varphi(x_2) d^4x_1 d^4x_2~,
\label{non-local}
\end{eqnarray}
where, $v(x_1, x_2)$ is a function of two spacelike points $x_1$ and $x_2$
and is common to other interactions in (\ref{L-varpsi-tilde}).
We assume that
$v(x, x) = 0$ and $v(x_1, x_2)$ can take non-zero values for $(x_1-x_2)^2 = O(\ell^2)$
($\ell$ is a fundamental length).
The vertex representing the non-local interaction is depicted in Fig. \ref{Fvertex}.
Here, the factor such as $4v(x_1, x_2)$ 
for $\lambda_{\varphi}^{(\varphi)}$ is omitted, for simplicity.
The same applies hereafter.

In the case that $v(x_1, x_2) = \delta^4(x_1-x_2+\xi)$ with $\xi^2 = O(\ell^2)$, 
the non-local interaction (\ref{non-local}) becomes
\begin{eqnarray}
-\int \lambda_{\varphi}^{(\varphi)}
\varphi^{\dagger}(x_1) \varphi(x_1)  \varphi^{\dagger}(x_1+\xi) \varphi(x_1+\xi) d^4x_1~.
\label{non-local-ex}
\end{eqnarray}
If $\xi^{\mu} = 0$, the interaction becomes local.
Then the self-interaction of $c_{\varphi}$ vanishes such that
$- \lambda_{\varphi}^{(c)} :c_{\varphi}^{\dagger} c_\varphi c_{\varphi}^{\dagger} c_\varphi:=0$
because of $c_{\varphi}^2 = 0$.
However, the self-interaction of $c_{\varphi}$ is indispensable,
because it is induced radiatively through the coupling between light and heavy fields
and it contains infinities that should be removed 
through the renormalization of relevant coupling constant.
This is the reason why we introduce non-local self-interactions.

$\tilde{\mathcal{L}}_{\rm T}$ has $OSp(2|2)$ invariance,
when the following relations among parameters hold
\begin{eqnarray}
M_{\varphi}^{(\varphi)2} = M_{\varphi}^{(c)2}~,~~
\lambda_{\varphi}^{(\varphi)} = \lambda_{\varphi}^{(\varphi,c)} = \lambda_{\varphi}^{(c)}~,~~
\lambda'^{(\varphi)} = \lambda'^{(c)}~.
\label{OSp-invL}
\end{eqnarray}

Let us study radiative corrections on parameters at the one-loop level,
without specifying the form of $v(x_1, x_2)$, and
examine whether $OSp(2|2)$ invariance holds at the quantum level.

First, we consider radiative corrections on $m_{\phi}^2$.
The one-loop diagrams concerning $\delta m_{\phi}^2$ are given in Fig. \ref{Fmphi}.
\begin{center}
\begin{figure}[hbtp]
\caption[mphi]{The one-loop diagrams of $\delta m_{\phi}^2$.}
\label{Fmphi}
\begin{center}
\unitlength 0.1in
\begin{picture}( 50.6000,  7.5000)(  2.1000, -9.1000)
%
\special{pn 13}%
\special{pa 210 800}%
\special{pa 1480 800}%
\special{fp}%
%
\special{pn 13}%
\special{ar 820 590 200 200  0.0000000 6.2831853}%
%
\special{pn 13}%
\special{pa 770 400}%
\special{pa 860 400}%
\special{fp}%
\special{sh 1}%
\special{pa 860 400}%
\special{pa 794 380}%
\special{pa 808 400}%
\special{pa 794 420}%
\special{pa 860 400}%
\special{fp}%
\put(7.5000,-3.3000){\makebox(0,0)[lb]{$\varphi$}}%
\put(6.9000,-10.7000){\makebox(0,0)[lb]{$\lambda'^{(\varphi)}$}}%
%
\special{pn 13}%
\special{pa 2110 800}%
\special{pa 3380 800}%
\special{fp}%
%
\special{pn 13}%
\special{ar 2720 590 200 200  0.0000000 6.2831853}%
%
\special{pn 13}%
\special{pa 2670 400}%
\special{pa 2760 400}%
\special{fp}%
\special{sh 1}%
\special{pa 2760 400}%
\special{pa 2694 380}%
\special{pa 2708 400}%
\special{pa 2694 420}%
\special{pa 2760 400}%
\special{fp}%
\put(26.4000,-3.3000){\makebox(0,0)[lb]{$c_{\varphi}$}}%
\put(26.5000,-10.7000){\makebox(0,0)[lb]{$\lambda'^{(c)}$}}%
%
\special{pn 13}%
\special{pa 4000 810}%
\special{pa 5270 810}%
\special{fp}%
%
\special{pn 13}%
\special{ar 4610 600 200 200  0.0000000 6.2831853}%
%
\special{pn 13}%
\special{pa 4560 410}%
\special{pa 4650 410}%
\special{fp}%
\special{sh 1}%
\special{pa 4650 410}%
\special{pa 4584 390}%
\special{pa 4598 410}%
\special{pa 4584 430}%
\special{pa 4650 410}%
\special{fp}%
\put(45.5000,-3.4000){\makebox(0,0)[lb]{$\phi$}}%
\put(45.5000,-10.8000){\makebox(0,0)[lb]{$\lambda_{\phi}$}}%
\put(17.3000,-8.4000){\makebox(0,0)[lb]{$+$}}%
\put(36.1000,-8.5000){\makebox(0,0)[lb]{$+$}}%
%
\special{pn 13}%
\special{pa 410 800}%
\special{pa 500 800}%
\special{fp}%
\special{sh 1}%
\special{pa 500 800}%
\special{pa 434 780}%
\special{pa 448 800}%
\special{pa 434 820}%
\special{pa 500 800}%
\special{fp}%
%
\special{pn 13}%
\special{pa 1150 800}%
\special{pa 1240 800}%
\special{fp}%
\special{sh 1}%
\special{pa 1240 800}%
\special{pa 1174 780}%
\special{pa 1188 800}%
\special{pa 1174 820}%
\special{pa 1240 800}%
\special{fp}%
%
\special{pn 13}%
\special{pa 2330 800}%
\special{pa 2420 800}%
\special{fp}%
\special{sh 1}%
\special{pa 2420 800}%
\special{pa 2354 780}%
\special{pa 2368 800}%
\special{pa 2354 820}%
\special{pa 2420 800}%
\special{fp}%
%
\special{pn 13}%
\special{pa 3060 800}%
\special{pa 3150 800}%
\special{fp}%
\special{sh 1}%
\special{pa 3150 800}%
\special{pa 3084 780}%
\special{pa 3098 800}%
\special{pa 3084 820}%
\special{pa 3150 800}%
\special{fp}%
%
\special{pn 13}%
\special{pa 4220 810}%
\special{pa 4310 810}%
\special{fp}%
\special{sh 1}%
\special{pa 4310 810}%
\special{pa 4244 790}%
\special{pa 4258 810}%
\special{pa 4244 830}%
\special{pa 4310 810}%
\special{fp}%
%
\special{pn 13}%
\special{pa 4920 810}%
\special{pa 5010 810}%
\special{fp}%
\special{sh 1}%
\special{pa 5010 810}%
\special{pa 4944 790}%
\special{pa 4958 810}%
\special{pa 4944 830}%
\special{pa 5010 810}%
\special{fp}%
\put(3.1000,-10.7000){\makebox(0,0)[lb]{$\phi^{\dagger}$}}%
\put(12.0000,-10.7000){\makebox(0,0)[lb]{$\phi$}}%
\end{picture}%
\end{center}
\end{figure}
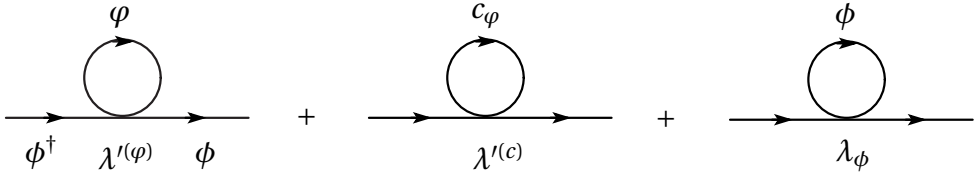
\end{center}
The contributions from the first two diagrams are canceled each other for the case with (\ref{OSp-invL}),
because the statistics of particles running in the loop is different from each other.
In this case, $\delta m_{\phi}^2$ is given by
\begin{eqnarray}
\delta m_{\phi}^2 
= -\frac{\lambda_{\phi}}{4\pi^2} m_{\phi}^2 \ln \frac{\Lambda^2}{m_{\phi}^2}~,
\label{deltamphi}
\end{eqnarray}
where we subtract the quadratic divergence.
In the same way, radiative corrections on $\lambda_{\phi}$
come from only the self-interaction of $\phi$, 
because the contributions from $\varphi$ and $c_{\varphi}$ are exactly canceled out.

Next, we study radiative corrections on $M_{\varphi}^{(\varphi)2}$ and $M_{\varphi}^{(c)2}$.
The one-loop diagrams of $\delta M_{\varphi}^{(\varphi)2}$ are given in Fig. \ref{FMvarphi}.
\begin{center}
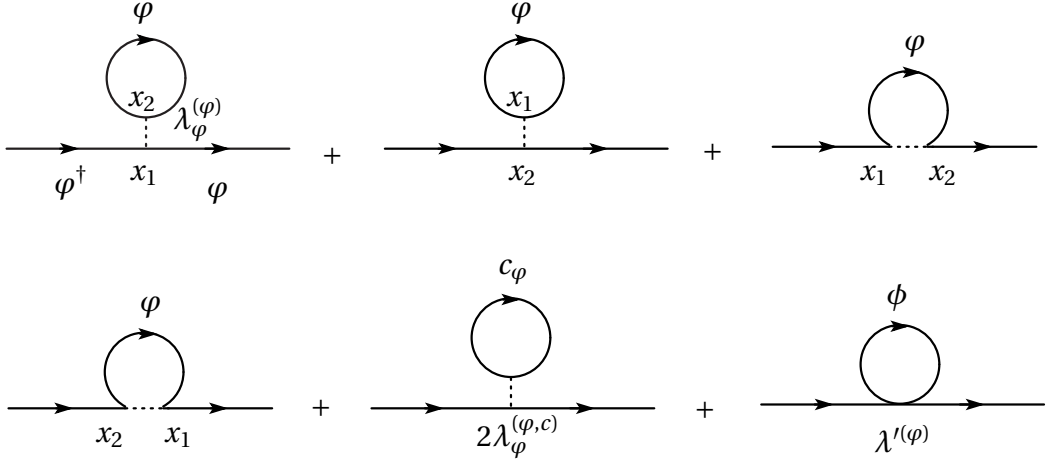
\begin{figure}[hbtp]
\caption[Mvarphi]{The one-loop diagrams of $\delta M_{\varphi}^{(\varphi)2}$.}
\label{FMvarphi}
\begin{center}
\unitlength 0.1in
\begin{picture}( 54.3000, 22.7000)(  1.2000,-24.3000)
%
\special{pn 13}%
\special{pa 120 980}%
\special{pa 1600 980}%
\special{fp}%
%
\special{pn 13}%
\special{ar 850 610 206 206  0.4182243 6.2831853}%
\special{ar 850 610 206 206  0.0000000 0.3985224}%
%
\special{pn 13}%
\special{pa 850 820}%
\special{pa 850 980}%
\special{dt 0.045}%
%
\special{pn 13}%
\special{pa 810 410}%
\special{pa 880 410}%
\special{fp}%
\special{sh 1}%
\special{pa 880 410}%
\special{pa 814 390}%
\special{pa 828 410}%
\special{pa 814 430}%
\special{pa 880 410}%
\special{fp}%
\put(7.8000,-3.3000){\makebox(0,0)[lb]{$\varphi$}}%
\put(3.7000,-12.5000){\makebox(0,0)[lb]{$\varphi^{\dagger}$}}%
\put(11.7000,-12.7000){\makebox(0,0)[lb]{$\varphi$}}%
\put(7.6000,-11.7000){\makebox(0,0)[lb]{$x_1$}}%
\put(7.5000,-7.8000){\makebox(0,0)[lb]{$x_2$}}%
%
\special{pn 13}%
\special{pa 1160 980}%
\special{pa 1270 980}%
\special{fp}%
\special{sh 1}%
\special{pa 1270 980}%
\special{pa 1204 960}%
\special{pa 1218 980}%
\special{pa 1204 1000}%
\special{pa 1270 980}%
\special{fp}%
%
\special{pn 13}%
\special{pa 400 980}%
\special{pa 480 980}%
\special{fp}%
\special{sh 1}%
\special{pa 480 980}%
\special{pa 414 960}%
\special{pa 428 980}%
\special{pa 414 1000}%
\special{pa 480 980}%
\special{fp}%
%
\special{pn 13}%
\special{pa 2100 980}%
\special{pa 3580 980}%
\special{fp}%
%
\special{pn 13}%
\special{ar 2830 610 206 206  0.4182243 6.2831853}%
\special{ar 2830 610 206 206  0.0000000 0.3985224}%
%
\special{pn 13}%
\special{pa 2830 820}%
\special{pa 2830 980}%
\special{dt 0.045}%
%
\special{pn 13}%
\special{pa 2790 410}%
\special{pa 2860 410}%
\special{fp}%
\special{sh 1}%
\special{pa 2860 410}%
\special{pa 2794 390}%
\special{pa 2808 410}%
\special{pa 2794 430}%
\special{pa 2860 410}%
\special{fp}%
\put(27.6000,-3.3000){\makebox(0,0)[lb]{$\varphi$}}%
\put(27.4000,-11.7000){\makebox(0,0)[lb]{$x_2$}}%
\put(27.3000,-7.8000){\makebox(0,0)[lb]{$x_1$}}%
%
\special{pn 13}%
\special{pa 3140 980}%
\special{pa 3250 980}%
\special{fp}%
\special{sh 1}%
\special{pa 3250 980}%
\special{pa 3184 960}%
\special{pa 3198 980}%
\special{pa 3184 1000}%
\special{pa 3250 980}%
\special{fp}%
%
\special{pn 13}%
\special{pa 2380 980}%
\special{pa 2460 980}%
\special{fp}%
\special{sh 1}%
\special{pa 2460 980}%
\special{pa 2394 960}%
\special{pa 2408 980}%
\special{pa 2394 1000}%
\special{pa 2460 980}%
\special{fp}%
%
\special{pn 13}%
\special{pa 2030 2340}%
\special{pa 3510 2340}%
\special{fp}%
%
\special{pn 13}%
\special{ar 2760 1970 206 206  0.4182243 6.2831853}%
\special{ar 2760 1970 206 206  0.0000000 0.3985224}%
%
\special{pn 13}%
\special{pa 2760 2180}%
\special{pa 2760 2340}%
\special{dt 0.045}%
%
\special{pn 13}%
\special{pa 2720 1770}%
\special{pa 2790 1770}%
\special{fp}%
\special{sh 1}%
\special{pa 2790 1770}%
\special{pa 2724 1750}%
\special{pa 2738 1770}%
\special{pa 2724 1790}%
\special{pa 2790 1770}%
\special{fp}%
\put(26.9000,-16.9000){\makebox(0,0)[lb]{$c_{\varphi}$}}%
%
\special{pn 13}%
\special{pa 3070 2340}%
\special{pa 3180 2340}%
\special{fp}%
\special{sh 1}%
\special{pa 3180 2340}%
\special{pa 3114 2320}%
\special{pa 3128 2340}%
\special{pa 3114 2360}%
\special{pa 3180 2340}%
\special{fp}%
%
\special{pn 13}%
\special{pa 2310 2340}%
\special{pa 2390 2340}%
\special{fp}%
\special{sh 1}%
\special{pa 2390 2340}%
\special{pa 2324 2320}%
\special{pa 2338 2340}%
\special{pa 2324 2360}%
\special{pa 2390 2340}%
\special{fp}%
\put(25.8000,-26.0000){\makebox(0,0)[lb]{$2\lambda_{\varphi}^{(\varphi,c)}$}}%
\put(9.9000,-9.4000){\makebox(0,0)[lb]{$\lambda_{\varphi}^{(\varphi)}$}}%
%
\special{pn 13}%
\special{pa 4070 2320}%
\special{pa 5550 2320}%
\special{fp}%
%
\special{pn 13}%
\special{ar 4796 2110 206 206  0.4182243 6.2831853}%
\special{ar 4796 2110 206 206  0.0000000 0.3985224}%
%
\special{pn 13}%
\special{pa 4756 1910}%
\special{pa 4826 1910}%
\special{fp}%
\special{sh 1}%
\special{pa 4826 1910}%
\special{pa 4760 1890}%
\special{pa 4774 1910}%
\special{pa 4760 1930}%
\special{pa 4826 1910}%
\special{fp}%
\put(47.2600,-18.3000){\makebox(0,0)[lb]{$\phi$}}%
%
\special{pn 13}%
\special{pa 5110 2320}%
\special{pa 5220 2320}%
\special{fp}%
\special{sh 1}%
\special{pa 5220 2320}%
\special{pa 5154 2300}%
\special{pa 5168 2320}%
\special{pa 5154 2340}%
\special{pa 5220 2320}%
\special{fp}%
%
\special{pn 13}%
\special{pa 4350 2320}%
\special{pa 4430 2320}%
\special{fp}%
\special{sh 1}%
\special{pa 4430 2320}%
\special{pa 4364 2300}%
\special{pa 4378 2320}%
\special{pa 4364 2340}%
\special{pa 4430 2320}%
\special{fp}%
\put(46.6000,-25.7000){\makebox(0,0)[lb]{$\lambda'^{(\varphi)}$}}%
%
\special{pn 13}%
\special{pa 4130 970}%
\special{pa 4740 970}%
\special{fp}%
%
\special{pn 13}%
\special{pa 4940 970}%
\special{pa 5510 970}%
\special{fp}%
%
\special{pn 13}%
\special{pa 4740 970}%
\special{pa 4970 970}%
\special{dt 0.045}%
%
\special{pn 13}%
\special{ar 4840 780 206 206  2.0736395 6.2831853}%
\special{ar 4840 780 206 206  0.0000000 1.0636978}%
%
\special{pn 13}%
\special{pa 4820 580}%
\special{pa 4890 580}%
\special{fp}%
\special{sh 1}%
\special{pa 4890 580}%
\special{pa 4824 560}%
\special{pa 4838 580}%
\special{pa 4824 600}%
\special{pa 4890 580}%
\special{fp}%
%
\special{pn 13}%
\special{pa 4380 970}%
\special{pa 4450 970}%
\special{fp}%
\special{sh 1}%
\special{pa 4450 970}%
\special{pa 4384 950}%
\special{pa 4398 970}%
\special{pa 4384 990}%
\special{pa 4450 970}%
\special{fp}%
%
\special{pn 13}%
\special{pa 5190 970}%
\special{pa 5260 970}%
\special{fp}%
\special{sh 1}%
\special{pa 5260 970}%
\special{pa 5194 950}%
\special{pa 5208 970}%
\special{pa 5194 990}%
\special{pa 5260 970}%
\special{fp}%
\put(49.4000,-11.7000){\makebox(0,0)[lb]{$x_2$}}%
\put(45.8000,-11.7000){\makebox(0,0)[lb]{$x_1$}}%
\put(48.2000,-5.0000){\makebox(0,0)[lb]{$\varphi$}}%
%
\special{pn 13}%
\special{pa 130 2340}%
\special{pa 740 2340}%
\special{fp}%
%
\special{pn 13}%
\special{pa 940 2340}%
\special{pa 1510 2340}%
\special{fp}%
%
\special{pn 13}%
\special{pa 740 2340}%
\special{pa 970 2340}%
\special{dt 0.045}%
%
\special{pn 13}%
\special{ar 840 2150 206 206  2.0736395 6.2831853}%
\special{ar 840 2150 206 206  0.0000000 1.0636978}%
%
\special{pn 13}%
\special{pa 820 1950}%
\special{pa 890 1950}%
\special{fp}%
\special{sh 1}%
\special{pa 890 1950}%
\special{pa 824 1930}%
\special{pa 838 1950}%
\special{pa 824 1970}%
\special{pa 890 1950}%
\special{fp}%
%
\special{pn 13}%
\special{pa 380 2340}%
\special{pa 450 2340}%
\special{fp}%
\special{sh 1}%
\special{pa 450 2340}%
\special{pa 384 2320}%
\special{pa 398 2340}%
\special{pa 384 2360}%
\special{pa 450 2340}%
\special{fp}%
%
\special{pn 13}%
\special{pa 1190 2340}%
\special{pa 1260 2340}%
\special{fp}%
\special{sh 1}%
\special{pa 1260 2340}%
\special{pa 1194 2320}%
\special{pa 1208 2340}%
\special{pa 1194 2360}%
\special{pa 1260 2340}%
\special{fp}%
\put(9.4000,-25.4000){\makebox(0,0)[lb]{$x_1$}}%
\put(5.8000,-25.4000){\makebox(0,0)[lb]{$x_2$}}%
\put(8.2000,-18.7000){\makebox(0,0)[lb]{$\varphi$}}%
\put(17.7000,-10.6000){\makebox(0,0)[lb]{$+$}}%
\put(37.6000,-10.4000){\makebox(0,0)[lb]{$+$}}%
\put(17.1000,-23.8000){\makebox(0,0)[lb]{$+$}}%
\put(37.2000,-23.9000){\makebox(0,0)[lb]{$+$}}%
\end{picture}%
\end{center}
\end{figure}
\end{center}
Those of $\delta M_{\varphi}^{(c)2}$ are given by exchanging $\varphi$ for $c_{\varphi}$,
$\lambda_{\varphi}^{(\varphi)}$ for $\lambda_{\varphi}^{(c)}$
and $\lambda'^{(\varphi)}$ for $\lambda'^{(c)}$.
The contributions from the first two diagrams and the fifth one are canceled
for the case with (\ref{OSp-invL}).
In this case, after the subtraction of quadratic divergence, $\delta M_{\varphi}^{(\varphi)2}$ is given by
\begin{eqnarray}
\delta M_{\varphi}^{(\varphi)2}
= -\frac{\lambda'^{(\varphi)}}{4\pi^2} m_{\phi}^2 \ln \frac{\Lambda^2}{m_{\phi}^2} 
+ 2 \lambda_{\varphi}^{(\varphi)} J_{\varphi}~,
\label{deltaMvarphi}
\end{eqnarray}
where $J_{\varphi}$ represents the sum of contributions from the third and fourth diagrams.
In the same way, $\delta M_{\varphi}^{(c)2}$ is given by
\begin{eqnarray}
\delta M_{\varphi}^{(c)2}
= -\frac{\lambda'^{(c)}}{4\pi^2} m_{\phi}^2 \ln \frac{\Lambda^2}{m_{\phi}^2} + 2 \lambda_{\varphi}^{(c)} J_{c_{\varphi}}~,
\label{deltaMc}
\end{eqnarray}
where $J_{c_{\varphi}}$ represents the counterpart to $J_{\varphi}$.
From (\ref{deltaMvarphi}) and (\ref{deltaMc}), we find that
$\delta M_{\varphi}^{(\varphi)2} = \delta M_{\varphi}^{(c)2}$ for the case with (\ref{OSp-invL})
and $J_{\varphi} = J_{c_{\varphi}}$
The equality $J_{\varphi} = J_{c_{\varphi}}$ holds if the following reasoning is correct.
In the third and fourth diagrams, $\varphi$ ($c_{\varphi}$) does not form the closed line by itself
(the loop is composed of $\varphi$ ($c_{\varphi}$) and the dashed line representing non-local interactions),
and hence an extra minus sign is not required for the propagation of $c_{\varphi}$.
In this case, the same size of contributions are expected for $J_{\varphi}$ and $J_{c_{\varphi}}$.

Finally, we study radiative corrections on $\lambda_{\varphi}^{(\varphi)}$, $\lambda_{\varphi}^{(c)}$
and $\lambda_{\varphi}^{(\varphi,c)}$.
The one-loop diagrams of $\delta(\lambda_{\varphi}^{(\varphi)}v(x_1, x_2))$ 
are given in Fig. \ref{Flambdavarphi}.
\begin{center}
\begin{figure}[hbtp]
\caption[lambdavarphi]{The one-loop diagrams of $\delta(\lambda_{\varphi}^{(\varphi)}v(x_1, x_2))$.}
\label{Flambdavarphi}
\begin{center}
\unitlength 0.1in
\begin{picture}( 52.5000, 26.2000)(  0.3000,-28.0000)
%
\special{pn 13}%
\special{pa 720 400}%
\special{pa 1010 810}%
\special{fp}%
%
\special{pn 13}%
\special{pa 1010 800}%
\special{pa 720 1210}%
\special{fp}%
%
\special{pn 13}%
\special{pa 1910 1210}%
\special{pa 1620 800}%
\special{fp}%
%
\special{pn 13}%
\special{pa 1620 810}%
\special{pa 1910 400}%
\special{fp}%
%
\special{pn 13}%
\special{ar 1320 800 198 198  0.7454195 6.2831853}%
\special{ar 1320 800 198 198  0.0000000 0.7140907}%
%
\special{pn 13}%
\special{pa 990 810}%
\special{pa 1110 810}%
\special{dt 0.045}%
%
\special{pn 13}%
\special{pa 1520 810}%
\special{pa 1630 810}%
\special{dt 0.045}%
%
\special{pn 13}%
\special{pa 1280 600}%
\special{pa 1350 600}%
\special{fp}%
\special{sh 1}%
\special{pa 1350 600}%
\special{pa 1284 580}%
\special{pa 1298 600}%
\special{pa 1284 620}%
\special{pa 1350 600}%
\special{fp}%
%
\special{pn 13}%
\special{pa 1360 1000}%
\special{pa 1270 1000}%
\special{fp}%
\special{sh 1}%
\special{pa 1270 1000}%
\special{pa 1338 1020}%
\special{pa 1324 1000}%
\special{pa 1338 980}%
\special{pa 1270 1000}%
\special{fp}%
%
\special{pn 13}%
\special{pa 890 640}%
\special{pa 840 560}%
\special{fp}%
\special{sh 1}%
\special{pa 840 560}%
\special{pa 858 628}%
\special{pa 868 606}%
\special{pa 892 606}%
\special{pa 840 560}%
\special{fp}%
%
\special{pn 13}%
\special{pa 1730 650}%
\special{pa 1770 580}%
\special{fp}%
\special{sh 1}%
\special{pa 1770 580}%
\special{pa 1720 628}%
\special{pa 1744 626}%
\special{pa 1754 648}%
\special{pa 1770 580}%
\special{fp}%
%
\special{pn 13}%
\special{pa 810 1110}%
\special{pa 850 1020}%
\special{fp}%
\special{sh 1}%
\special{pa 850 1020}%
\special{pa 806 1074}%
\special{pa 828 1070}%
\special{pa 842 1090}%
\special{pa 850 1020}%
\special{fp}%
\put(6.1000,-3.6000){\makebox(0,0)[lb]{$\varphi$}}%
\put(19.0000,-3.5000){\makebox(0,0)[lb]{$\varphi$}}%
\put(6.2000,-14.1000){\makebox(0,0)[lb]{$\varphi^{\dagger}$}}%
\put(19.1000,-14.1000){\makebox(0,0)[lb]{$\varphi^{\dagger}$}}%
\put(12.6000,-8.7000){\makebox(0,0)[lb]{$\varphi$}}%
%
\special{pn 13}%
\special{pa 1810 1070}%
\special{pa 1730 980}%
\special{fp}%
\special{sh 1}%
\special{pa 1730 980}%
\special{pa 1760 1044}%
\special{pa 1766 1020}%
\special{pa 1790 1018}%
\special{pa 1730 980}%
\special{fp}%
\put(17.2000,-9.3000){\makebox(0,0)[lb]{$\lambda_{\varphi}^{(\varphi)}$}}%
\put(6.5000,-9.3000){\makebox(0,0)[lb]{$\lambda_{\varphi}^{(\varphi)}$}}%
%
\special{pn 13}%
\special{pa 2370 1860}%
\special{pa 2660 2270}%
\special{fp}%
%
\special{pn 13}%
\special{pa 2660 2260}%
\special{pa 2370 2670}%
\special{fp}%
%
\special{pn 13}%
\special{pa 3560 2670}%
\special{pa 3270 2260}%
\special{fp}%
%
\special{pn 13}%
\special{pa 3270 2270}%
\special{pa 3560 1860}%
\special{fp}%
%
\special{pn 13}%
\special{ar 2970 2260 198 198  0.7454195 6.2831853}%
\special{ar 2970 2260 198 198  0.0000000 0.7140907}%
%
\special{pn 13}%
\special{pa 2640 2270}%
\special{pa 2760 2270}%
\special{dt 0.045}%
%
\special{pn 13}%
\special{pa 3170 2270}%
\special{pa 3280 2270}%
\special{dt 0.045}%
%
\special{pn 13}%
\special{pa 2930 2060}%
\special{pa 3000 2060}%
\special{fp}%
\special{sh 1}%
\special{pa 3000 2060}%
\special{pa 2934 2040}%
\special{pa 2948 2060}%
\special{pa 2934 2080}%
\special{pa 3000 2060}%
\special{fp}%
%
\special{pn 13}%
\special{pa 3010 2460}%
\special{pa 2920 2460}%
\special{fp}%
\special{sh 1}%
\special{pa 2920 2460}%
\special{pa 2988 2480}%
\special{pa 2974 2460}%
\special{pa 2988 2440}%
\special{pa 2920 2460}%
\special{fp}%
%
\special{pn 13}%
\special{pa 2540 2100}%
\special{pa 2490 2020}%
\special{fp}%
\special{sh 1}%
\special{pa 2490 2020}%
\special{pa 2508 2088}%
\special{pa 2518 2066}%
\special{pa 2542 2066}%
\special{pa 2490 2020}%
\special{fp}%
%
\special{pn 13}%
\special{pa 3380 2110}%
\special{pa 3420 2040}%
\special{fp}%
\special{sh 1}%
\special{pa 3420 2040}%
\special{pa 3370 2088}%
\special{pa 3394 2086}%
\special{pa 3404 2108}%
\special{pa 3420 2040}%
\special{fp}%
%
\special{pn 13}%
\special{pa 2450 2570}%
\special{pa 2490 2480}%
\special{fp}%
\special{sh 1}%
\special{pa 2490 2480}%
\special{pa 2446 2534}%
\special{pa 2468 2530}%
\special{pa 2482 2550}%
\special{pa 2490 2480}%
\special{fp}%
\put(29.0000,-23.4000){\makebox(0,0)[lb]{$c_{\varphi}$}}%
%
\special{pn 13}%
\special{pa 3460 2530}%
\special{pa 3380 2440}%
\special{fp}%
\special{sh 1}%
\special{pa 3380 2440}%
\special{pa 3410 2504}%
\special{pa 3416 2480}%
\special{pa 3440 2478}%
\special{pa 3380 2440}%
\special{fp}%
\put(33.7000,-23.8000){\makebox(0,0)[lb]{$2\lambda_{\varphi}^{(\varphi,c)}$}}%
\put(21.0000,-23.6000){\makebox(0,0)[lb]{$2\lambda_{\varphi}^{(\varphi,c)}$}}%
\put(0.3000,-9.1000){\makebox(0,0)[lb]{$4 \times$}}%
%
\special{pn 13}%
\special{pa 4170 1880}%
\special{pa 4460 2290}%
\special{fp}%
%
\special{pn 13}%
\special{pa 4460 2280}%
\special{pa 4170 2690}%
\special{fp}%
%
\special{pn 13}%
\special{pa 3440 1150}%
\special{pa 3230 880}%
\special{fp}%
%
\special{pn 13}%
\special{pa 3220 720}%
\special{pa 3450 410}%
\special{fp}%
%
\special{pn 13}%
\special{pa 3010 600}%
\special{pa 3080 600}%
\special{fp}%
\special{sh 1}%
\special{pa 3080 600}%
\special{pa 3014 580}%
\special{pa 3028 600}%
\special{pa 3014 620}%
\special{pa 3080 600}%
\special{fp}%
%
\special{pn 13}%
\special{pa 3070 1010}%
\special{pa 2980 1010}%
\special{fp}%
\special{sh 1}%
\special{pa 2980 1010}%
\special{pa 3048 1030}%
\special{pa 3034 1010}%
\special{pa 3048 990}%
\special{pa 2980 1010}%
\special{fp}%
%
\special{pn 13}%
\special{pa 4340 2120}%
\special{pa 4290 2040}%
\special{fp}%
\special{sh 1}%
\special{pa 4290 2040}%
\special{pa 4308 2108}%
\special{pa 4318 2086}%
\special{pa 4342 2086}%
\special{pa 4290 2040}%
\special{fp}%
%
\special{pn 13}%
\special{pa 4260 2570}%
\special{pa 4300 2480}%
\special{fp}%
\special{sh 1}%
\special{pa 4300 2480}%
\special{pa 4256 2534}%
\special{pa 4278 2530}%
\special{pa 4292 2550}%
\special{pa 4300 2480}%
\special{fp}%
\put(23.4000,-9.0000){\makebox(0,0)[lb]{$\lambda_{\varphi}^{(\varphi)}$}}%
%
\special{pn 13}%
\special{pa 2420 410}%
\special{pa 2710 820}%
\special{fp}%
%
\special{pn 13}%
\special{pa 2710 810}%
\special{pa 2420 1220}%
\special{fp}%
%
\special{pn 13}%
\special{pa 2690 820}%
\special{pa 2810 820}%
\special{dt 0.045}%
%
\special{pn 13}%
\special{pa 2590 650}%
\special{pa 2540 570}%
\special{fp}%
\special{sh 1}%
\special{pa 2540 570}%
\special{pa 2558 638}%
\special{pa 2568 616}%
\special{pa 2592 616}%
\special{pa 2540 570}%
\special{fp}%
%
\special{pn 13}%
\special{pa 2530 1080}%
\special{pa 2570 990}%
\special{fp}%
\special{sh 1}%
\special{pa 2570 990}%
\special{pa 2526 1044}%
\special{pa 2548 1040}%
\special{pa 2562 1060}%
\special{pa 2570 990}%
\special{fp}%
%
\special{pn 13}%
\special{pa 3230 730}%
\special{pa 3230 890}%
\special{dt 0.045}%
%
\special{pn 13}%
\special{ar 3030 810 206 206  0.3529904 5.8408111}%
%
\special{pn 13}%
\special{pa 3370 1060}%
\special{pa 3310 970}%
\special{fp}%
\special{sh 1}%
\special{pa 3310 970}%
\special{pa 3330 1038}%
\special{pa 3340 1014}%
\special{pa 3364 1014}%
\special{pa 3310 970}%
\special{fp}%
%
\special{pn 13}%
\special{pa 3320 580}%
\special{pa 3390 510}%
\special{fp}%
\special{sh 1}%
\special{pa 3390 510}%
\special{pa 3330 544}%
\special{pa 3352 548}%
\special{pa 3358 572}%
\special{pa 3390 510}%
\special{fp}%
%
\special{pn 13}%
\special{pa 3890 410}%
\special{pa 4162 714}%
\special{fp}%
%
\special{pn 13}%
\special{pa 4162 896}%
\special{pa 3934 1208}%
\special{fp}%
%
\special{pn 13}%
\special{pa 4380 600}%
\special{pa 4310 600}%
\special{fp}%
\special{sh 1}%
\special{pa 4310 600}%
\special{pa 4378 620}%
\special{pa 4364 600}%
\special{pa 4378 580}%
\special{pa 4310 600}%
\special{fp}%
%
\special{pn 13}%
\special{pa 4320 1010}%
\special{pa 4410 1010}%
\special{fp}%
\special{sh 1}%
\special{pa 4410 1010}%
\special{pa 4344 990}%
\special{pa 4358 1010}%
\special{pa 4344 1030}%
\special{pa 4410 1010}%
\special{fp}%
%
\special{pn 13}%
\special{pa 4964 1200}%
\special{pa 4672 792}%
\special{fp}%
%
\special{pn 13}%
\special{pa 4672 802}%
\special{pa 4958 390}%
\special{fp}%
%
\special{pn 13}%
\special{pa 4692 792}%
\special{pa 4572 792}%
\special{dt 0.045}%
%
\special{pn 13}%
\special{pa 4152 886}%
\special{pa 4150 726}%
\special{dt 0.045}%
%
\special{pn 13}%
\special{ar 4352 804 206 206  3.4882602 6.2831853}%
\special{ar 4352 804 206 206  0.0000000 2.6949285}%
%
\special{pn 13}%
\special{pa 4020 1100}%
\special{pa 4090 1010}%
\special{fp}%
\special{sh 1}%
\special{pa 4090 1010}%
\special{pa 4034 1050}%
\special{pa 4058 1052}%
\special{pa 4066 1076}%
\special{pa 4090 1010}%
\special{fp}%
%
\special{pn 13}%
\special{pa 4022 574}%
\special{pa 3962 494}%
\special{fp}%
\special{sh 1}%
\special{pa 3962 494}%
\special{pa 3986 558}%
\special{pa 3994 536}%
\special{pa 4018 534}%
\special{pa 3962 494}%
\special{fp}%
%
\special{pn 13}%
\special{pa 4850 1050}%
\special{pa 4800 960}%
\special{fp}%
\special{sh 1}%
\special{pa 4800 960}%
\special{pa 4816 1028}%
\special{pa 4826 1008}%
\special{pa 4850 1010}%
\special{pa 4800 960}%
\special{fp}%
\put(33.3000,-8.8000){\makebox(0,0)[lb]{$\lambda_{\varphi}^{(\varphi)}$}}%
%
\special{pn 13}%
\special{pa 5280 2680}%
\special{pa 4990 2270}%
\special{fp}%
%
\special{pn 13}%
\special{pa 4990 2280}%
\special{pa 5280 1870}%
\special{fp}%
%
\special{pn 13}%
\special{pa 5140 2080}%
\special{pa 5180 2010}%
\special{fp}%
\special{sh 1}%
\special{pa 5180 2010}%
\special{pa 5130 2058}%
\special{pa 5154 2056}%
\special{pa 5164 2078}%
\special{pa 5180 2010}%
\special{fp}%
%
\special{pn 13}%
\special{pa 5190 2540}%
\special{pa 5110 2450}%
\special{fp}%
\special{sh 1}%
\special{pa 5110 2450}%
\special{pa 5140 2514}%
\special{pa 5146 2490}%
\special{pa 5170 2488}%
\special{pa 5110 2450}%
\special{fp}%
%
\special{pn 13}%
\special{ar 4730 2280 256 256  0.0000000 6.2831853}%
%
\special{pn 13}%
\special{pa 4690 2030}%
\special{pa 4760 2030}%
\special{fp}%
\special{sh 1}%
\special{pa 4760 2030}%
\special{pa 4694 2010}%
\special{pa 4708 2030}%
\special{pa 4694 2050}%
\special{pa 4760 2030}%
\special{fp}%
%
\special{pn 13}%
\special{pa 4770 2530}%
\special{pa 4680 2530}%
\special{fp}%
\special{sh 1}%
\special{pa 4680 2530}%
\special{pa 4748 2550}%
\special{pa 4734 2530}%
\special{pa 4748 2510}%
\special{pa 4680 2530}%
\special{fp}%
\put(46.7000,-23.4000){\makebox(0,0)[lb]{$\phi$}}%
\put(40.7000,-23.3000){\makebox(0,0)[lb]{$\lambda'^{(\varphi)}$}}%
\put(51.0000,-23.3000){\makebox(0,0)[lb]{$\lambda'^{(\varphi)}$}}%
%
\special{pn 13}%
\special{ar 830 2210 184 184  3.4253868 5.8925783}%
%
\special{pn 13}%
\special{ar 828 2272 184 184  0.4124104 2.8706458}%
%
\special{pn 13}%
\special{pa 870 2030}%
\special{pa 780 2030}%
\special{fp}%
\special{sh 1}%
\special{pa 780 2030}%
\special{pa 848 2050}%
\special{pa 834 2030}%
\special{pa 848 2010}%
\special{pa 780 2030}%
\special{fp}%
%
\special{pn 13}%
\special{pa 798 2452}%
\special{pa 868 2452}%
\special{fp}%
\special{sh 1}%
\special{pa 868 2452}%
\special{pa 800 2432}%
\special{pa 814 2452}%
\special{pa 800 2472}%
\special{pa 868 2452}%
\special{fp}%
%
\special{pn 13}%
\special{pa 660 2150}%
\special{pa 660 2340}%
\special{dt 0.045}%
%
\special{pn 13}%
\special{pa 1000 2140}%
\special{pa 1000 2330}%
\special{dt 0.045}%
%
\special{pn 13}%
\special{pa 650 2130}%
\special{pa 260 1870}%
\special{fp}%
%
\special{pn 13}%
\special{pa 650 2340}%
\special{pa 260 2590}%
\special{fp}%
%
\special{pn 13}%
\special{pa 1000 2140}%
\special{pa 1420 2600}%
\special{fp}%
%
\special{pn 13}%
\special{pa 1120 2200}%
\special{pa 1400 1880}%
\special{fp}%
%
\special{pn 13}%
\special{pa 1070 2270}%
\special{pa 1000 2350}%
\special{fp}%
%
\special{pn 13}%
\special{pa 420 2490}%
\special{pa 500 2440}%
\special{fp}%
\special{sh 1}%
\special{pa 500 2440}%
\special{pa 434 2458}%
\special{pa 456 2468}%
\special{pa 454 2492}%
\special{pa 500 2440}%
\special{fp}%
%
\special{pn 13}%
\special{pa 520 2030}%
\special{pa 410 1980}%
\special{fp}%
\special{sh 1}%
\special{pa 410 1980}%
\special{pa 462 2026}%
\special{pa 460 2002}%
\special{pa 480 1990}%
\special{pa 410 1980}%
\special{fp}%
%
\special{pn 13}%
\special{pa 1290 2460}%
\special{pa 1220 2370}%
\special{fp}%
\special{sh 1}%
\special{pa 1220 2370}%
\special{pa 1246 2436}%
\special{pa 1254 2412}%
\special{pa 1278 2410}%
\special{pa 1220 2370}%
\special{fp}%
%
\special{pn 13}%
\special{pa 1220 2080}%
\special{pa 1280 2010}%
\special{fp}%
\special{sh 1}%
\special{pa 1280 2010}%
\special{pa 1222 2048}%
\special{pa 1246 2050}%
\special{pa 1252 2074}%
\special{pa 1280 2010}%
\special{fp}%
%
\special{pn 8}%
\special{pa 410 300}%
\special{pa 260 400}%
\special{fp}%
%
\special{pn 8}%
\special{pa 270 390}%
\special{pa 270 1320}%
\special{fp}%
%
\special{pn 8}%
\special{pa 1600 2800}%
\special{pa 1720 2710}%
\special{fp}%
%
\special{pn 8}%
\special{pa 1710 2720}%
\special{pa 1710 1790}%
\special{fp}%
%
\special{pn 8}%
\special{pa 1710 1790}%
\special{pa 1550 1700}%
\special{fp}%
%
\special{pn 8}%
\special{pa 270 1310}%
\special{pa 420 1390}%
\special{fp}%
\put(20.7000,-8.9000){\makebox(0,0)[lb]{$+$}}%
\put(37.2000,-9.2000){\makebox(0,0)[lb]{$+$}}%
\put(18.9000,-23.1000){\makebox(0,0)[lb]{$+$}}%
\put(38.5000,-23.1000){\makebox(0,0)[lb]{$+$}}%
%
\special{pn 13}%
\special{pa 4800 610}%
\special{pa 4870 530}%
\special{fp}%
\special{sh 1}%
\special{pa 4870 530}%
\special{pa 4812 568}%
\special{pa 4836 570}%
\special{pa 4842 594}%
\special{pa 4870 530}%
\special{fp}%
\end{picture}%
\end{center}
\end{figure}
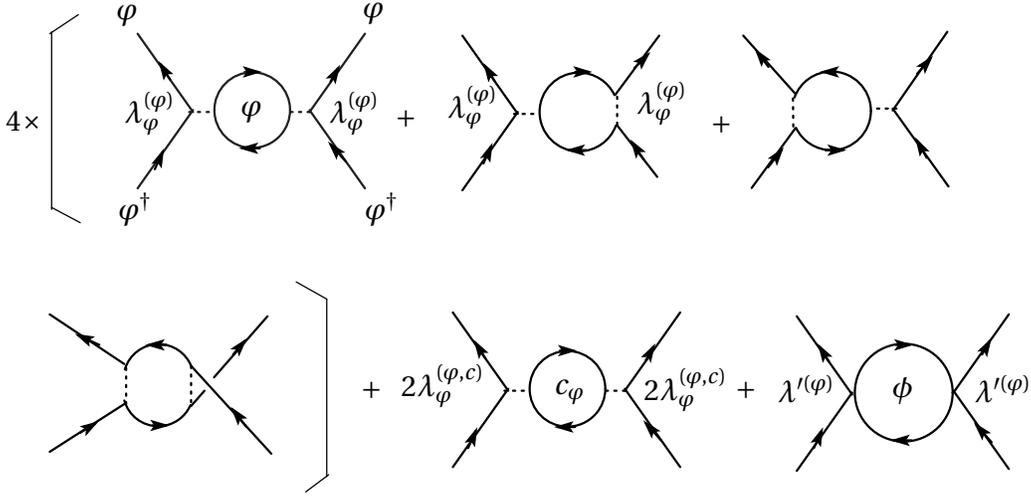
\end{center}
Here, the factor $4 \times$ comes from the fact that there are four ways
to contract external lines with non-local interaction points.
Those of $\delta(\lambda_{\varphi}^{(c)}v(x_1, x_2))$
are given by exchanging $\varphi$ for $c_{\varphi}$,
$\lambda_{\varphi}^{(\varphi)}$ for $\lambda_{\varphi}^{(c)}$
and $\lambda'^{(\varphi)}$ for $\lambda'^{(c)}$.
The contributions from the first and fifth diagrams are canceled 
for the case with (\ref{OSp-invL}).

The one-loop diagrams of $\delta(\lambda_{\varphi}^{(\varphi,c)}v(x_1, x_2))$ 
are given in Fig. \ref{Flambdavarphic}.
\begin{center}
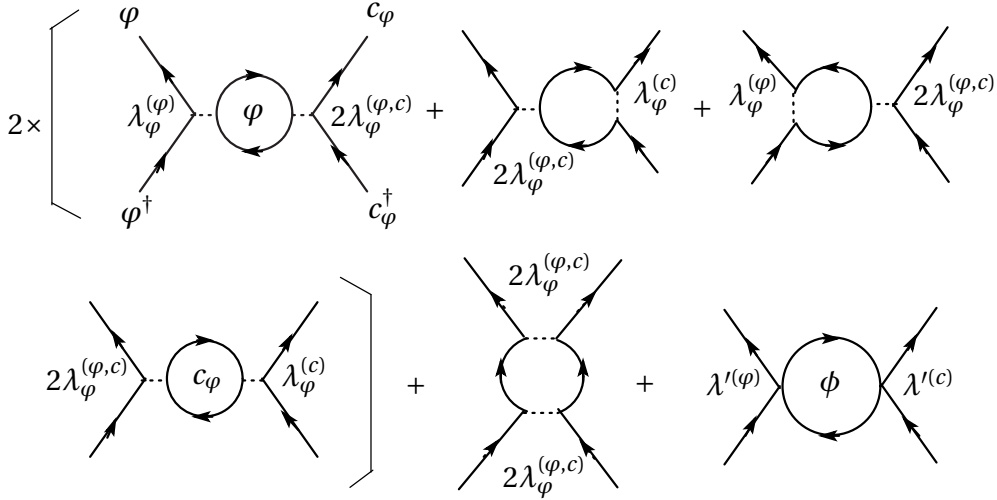
\begin{figure}[hbtp]
\caption[lambdavarphi]{The one-loop diagrams of 
$\delta(\lambda_{\varphi}^{(\varphi,c)} v(x_1, x_2))$.}
\label{Flambdavarphic}
\begin{center}
\unitlength 0.1in
\begin{picture}( 49.2300, 25.9000)(  0.4000,-28.1000)
%
\special{pn 13}%
\special{pa 730 440}%
\special{pa 1020 850}%
\special{fp}%
%
\special{pn 13}%
\special{pa 1020 840}%
\special{pa 730 1250}%
\special{fp}%
%
\special{pn 13}%
\special{pa 1920 1250}%
\special{pa 1630 840}%
\special{fp}%
%
\special{pn 13}%
\special{pa 1630 850}%
\special{pa 1920 440}%
\special{fp}%
%
\special{pn 13}%
\special{ar 1330 840 198 198  0.7454195 6.2831853}%
\special{ar 1330 840 198 198  0.0000000 0.7140907}%
%
\special{pn 13}%
\special{pa 1000 850}%
\special{pa 1120 850}%
\special{dt 0.045}%
%
\special{pn 13}%
\special{pa 1530 850}%
\special{pa 1640 850}%
\special{dt 0.045}%
%
\special{pn 13}%
\special{pa 1290 640}%
\special{pa 1360 640}%
\special{fp}%
\special{sh 1}%
\special{pa 1360 640}%
\special{pa 1294 620}%
\special{pa 1308 640}%
\special{pa 1294 660}%
\special{pa 1360 640}%
\special{fp}%
%
\special{pn 13}%
\special{pa 1370 1040}%
\special{pa 1280 1040}%
\special{fp}%
\special{sh 1}%
\special{pa 1280 1040}%
\special{pa 1348 1060}%
\special{pa 1334 1040}%
\special{pa 1348 1020}%
\special{pa 1280 1040}%
\special{fp}%
%
\special{pn 13}%
\special{pa 900 680}%
\special{pa 850 600}%
\special{fp}%
\special{sh 1}%
\special{pa 850 600}%
\special{pa 868 668}%
\special{pa 878 646}%
\special{pa 902 646}%
\special{pa 850 600}%
\special{fp}%
%
\special{pn 13}%
\special{pa 1740 690}%
\special{pa 1780 620}%
\special{fp}%
\special{sh 1}%
\special{pa 1780 620}%
\special{pa 1730 668}%
\special{pa 1754 666}%
\special{pa 1764 688}%
\special{pa 1780 620}%
\special{fp}%
%
\special{pn 13}%
\special{pa 820 1150}%
\special{pa 860 1060}%
\special{fp}%
\special{sh 1}%
\special{pa 860 1060}%
\special{pa 816 1114}%
\special{pa 838 1110}%
\special{pa 852 1130}%
\special{pa 860 1060}%
\special{fp}%
\put(6.2000,-4.0000){\makebox(0,0)[lb]{$\varphi$}}%
\put(19.1000,-3.9000){\makebox(0,0)[lb]{$c_{\varphi}$}}%
\put(6.3000,-14.5000){\makebox(0,0)[lb]{$\varphi^{\dagger}$}}%
\put(19.2000,-14.5000){\makebox(0,0)[lb]{$c_{\varphi}^{\dagger}$}}%
\put(12.7000,-9.1000){\makebox(0,0)[lb]{$\varphi$}}%
%
\special{pn 13}%
\special{pa 1820 1110}%
\special{pa 1740 1020}%
\special{fp}%
\special{sh 1}%
\special{pa 1740 1020}%
\special{pa 1770 1084}%
\special{pa 1776 1060}%
\special{pa 1800 1058}%
\special{pa 1740 1020}%
\special{fp}%
\put(17.3000,-9.7000){\makebox(0,0)[lb]{$2\lambda_{\varphi}^{(\varphi,c)}$}}%
\put(6.6000,-9.7000){\makebox(0,0)[lb]{$\lambda_{\varphi}^{(\varphi)}$}}%
%
\special{pn 13}%
\special{pa 470 1830}%
\special{pa 760 2240}%
\special{fp}%
%
\special{pn 13}%
\special{pa 760 2230}%
\special{pa 470 2640}%
\special{fp}%
%
\special{pn 13}%
\special{pa 1660 2640}%
\special{pa 1370 2230}%
\special{fp}%
%
\special{pn 13}%
\special{pa 1370 2240}%
\special{pa 1660 1830}%
\special{fp}%
%
\special{pn 13}%
\special{ar 1070 2230 198 198  0.7454195 6.2831853}%
\special{ar 1070 2230 198 198  0.0000000 0.7140907}%
%
\special{pn 13}%
\special{pa 740 2240}%
\special{pa 860 2240}%
\special{dt 0.045}%
%
\special{pn 13}%
\special{pa 1270 2240}%
\special{pa 1380 2240}%
\special{dt 0.045}%
%
\special{pn 13}%
\special{pa 1030 2030}%
\special{pa 1100 2030}%
\special{fp}%
\special{sh 1}%
\special{pa 1100 2030}%
\special{pa 1034 2010}%
\special{pa 1048 2030}%
\special{pa 1034 2050}%
\special{pa 1100 2030}%
\special{fp}%
%
\special{pn 13}%
\special{pa 1110 2430}%
\special{pa 1020 2430}%
\special{fp}%
\special{sh 1}%
\special{pa 1020 2430}%
\special{pa 1088 2450}%
\special{pa 1074 2430}%
\special{pa 1088 2410}%
\special{pa 1020 2430}%
\special{fp}%
%
\special{pn 13}%
\special{pa 640 2070}%
\special{pa 590 1990}%
\special{fp}%
\special{sh 1}%
\special{pa 590 1990}%
\special{pa 608 2058}%
\special{pa 618 2036}%
\special{pa 642 2036}%
\special{pa 590 1990}%
\special{fp}%
%
\special{pn 13}%
\special{pa 1480 2080}%
\special{pa 1520 2010}%
\special{fp}%
\special{sh 1}%
\special{pa 1520 2010}%
\special{pa 1470 2058}%
\special{pa 1494 2056}%
\special{pa 1504 2078}%
\special{pa 1520 2010}%
\special{fp}%
%
\special{pn 13}%
\special{pa 560 2540}%
\special{pa 600 2450}%
\special{fp}%
\special{sh 1}%
\special{pa 600 2450}%
\special{pa 556 2504}%
\special{pa 578 2500}%
\special{pa 592 2520}%
\special{pa 600 2450}%
\special{fp}%
\put(10.0000,-23.1000){\makebox(0,0)[lb]{$c_{\varphi}$}}%
%
\special{pn 13}%
\special{pa 1560 2500}%
\special{pa 1480 2410}%
\special{fp}%
\special{sh 1}%
\special{pa 1480 2410}%
\special{pa 1510 2474}%
\special{pa 1516 2450}%
\special{pa 1540 2448}%
\special{pa 1480 2410}%
\special{fp}%
\put(14.7000,-23.3000){\makebox(0,0)[lb]{$\lambda_{\varphi}^{(c)}$}}%
\put(2.4000,-23.5000){\makebox(0,0)[lb]{$2\lambda_{\varphi}^{(\varphi,c)}$}}%
\put(0.4000,-9.5000){\makebox(0,0)[lb]{$2 \times$}}%
%
\special{pn 13}%
\special{pa 3790 1870}%
\special{pa 4080 2280}%
\special{fp}%
%
\special{pn 13}%
\special{pa 4080 2270}%
\special{pa 3790 2680}%
\special{fp}%
%
\special{pn 13}%
\special{pa 3440 1150}%
\special{pa 3230 880}%
\special{fp}%
%
\special{pn 13}%
\special{pa 3220 720}%
\special{pa 3450 410}%
\special{fp}%
%
\special{pn 13}%
\special{pa 3010 600}%
\special{pa 3080 600}%
\special{fp}%
\special{sh 1}%
\special{pa 3080 600}%
\special{pa 3014 580}%
\special{pa 3028 600}%
\special{pa 3014 620}%
\special{pa 3080 600}%
\special{fp}%
%
\special{pn 13}%
\special{pa 3070 1010}%
\special{pa 2980 1010}%
\special{fp}%
\special{sh 1}%
\special{pa 2980 1010}%
\special{pa 3048 1030}%
\special{pa 3034 1010}%
\special{pa 3048 990}%
\special{pa 2980 1010}%
\special{fp}%
%
\special{pn 13}%
\special{pa 3960 2110}%
\special{pa 3910 2030}%
\special{fp}%
\special{sh 1}%
\special{pa 3910 2030}%
\special{pa 3928 2098}%
\special{pa 3938 2076}%
\special{pa 3962 2076}%
\special{pa 3910 2030}%
\special{fp}%
\put(25.8000,-12.7000){\makebox(0,0)[lb]{$2\lambda_{\varphi}^{(\varphi,c)}$}}%
%
\special{pn 13}%
\special{pa 2420 410}%
\special{pa 2710 820}%
\special{fp}%
%
\special{pn 13}%
\special{pa 2710 810}%
\special{pa 2420 1220}%
\special{fp}%
%
\special{pn 13}%
\special{pa 2690 820}%
\special{pa 2810 820}%
\special{dt 0.045}%
%
\special{pn 13}%
\special{pa 2590 650}%
\special{pa 2540 570}%
\special{fp}%
\special{sh 1}%
\special{pa 2540 570}%
\special{pa 2558 638}%
\special{pa 2568 616}%
\special{pa 2592 616}%
\special{pa 2540 570}%
\special{fp}%
%
\special{pn 13}%
\special{pa 2530 1080}%
\special{pa 2570 990}%
\special{fp}%
\special{sh 1}%
\special{pa 2570 990}%
\special{pa 2526 1044}%
\special{pa 2548 1040}%
\special{pa 2562 1060}%
\special{pa 2570 990}%
\special{fp}%
%
\special{pn 13}%
\special{pa 3230 730}%
\special{pa 3230 890}%
\special{dt 0.045}%
%
\special{pn 13}%
\special{ar 3030 810 206 206  0.3529904 5.8408111}%
%
\special{pn 13}%
\special{pa 3370 1060}%
\special{pa 3310 970}%
\special{fp}%
\special{sh 1}%
\special{pa 3310 970}%
\special{pa 3330 1038}%
\special{pa 3340 1014}%
\special{pa 3364 1014}%
\special{pa 3310 970}%
\special{fp}%
%
\special{pn 13}%
\special{pa 3320 580}%
\special{pa 3390 510}%
\special{fp}%
\special{sh 1}%
\special{pa 3390 510}%
\special{pa 3330 544}%
\special{pa 3352 548}%
\special{pa 3358 572}%
\special{pa 3390 510}%
\special{fp}%
%
\special{pn 13}%
\special{pa 3890 410}%
\special{pa 4162 714}%
\special{fp}%
%
\special{pn 13}%
\special{pa 4162 896}%
\special{pa 3934 1208}%
\special{fp}%
%
\special{pn 13}%
\special{pa 4380 600}%
\special{pa 4310 600}%
\special{fp}%
\special{sh 1}%
\special{pa 4310 600}%
\special{pa 4378 620}%
\special{pa 4364 600}%
\special{pa 4378 580}%
\special{pa 4310 600}%
\special{fp}%
%
\special{pn 13}%
\special{pa 4320 1010}%
\special{pa 4410 1010}%
\special{fp}%
\special{sh 1}%
\special{pa 4410 1010}%
\special{pa 4344 990}%
\special{pa 4358 1010}%
\special{pa 4344 1030}%
\special{pa 4410 1010}%
\special{fp}%
%
\special{pn 13}%
\special{pa 4964 1200}%
\special{pa 4672 792}%
\special{fp}%
%
\special{pn 13}%
\special{pa 4672 802}%
\special{pa 4958 390}%
\special{fp}%
%
\special{pn 13}%
\special{pa 4692 792}%
\special{pa 4572 792}%
\special{dt 0.045}%
%
\special{pn 13}%
\special{pa 4152 886}%
\special{pa 4150 726}%
\special{dt 0.045}%
%
\special{pn 13}%
\special{ar 4352 804 206 206  3.4882602 6.2831853}%
\special{ar 4352 804 206 206  0.0000000 2.6949285}%
%
\special{pn 13}%
\special{pa 4010 1100}%
\special{pa 4080 1010}%
\special{fp}%
\special{sh 1}%
\special{pa 4080 1010}%
\special{pa 4024 1050}%
\special{pa 4048 1052}%
\special{pa 4056 1076}%
\special{pa 4080 1010}%
\special{fp}%
%
\special{pn 13}%
\special{pa 4022 574}%
\special{pa 3962 494}%
\special{fp}%
\special{sh 1}%
\special{pa 3962 494}%
\special{pa 3986 558}%
\special{pa 3994 536}%
\special{pa 4018 534}%
\special{pa 3962 494}%
\special{fp}%
%
\special{pn 13}%
\special{pa 4850 1050}%
\special{pa 4800 960}%
\special{fp}%
\special{sh 1}%
\special{pa 4800 960}%
\special{pa 4816 1028}%
\special{pa 4826 1008}%
\special{pa 4850 1010}%
\special{pa 4800 960}%
\special{fp}%
%
\special{pn 13}%
\special{pa 4800 620}%
\special{pa 4890 500}%
\special{fp}%
\special{sh 1}%
\special{pa 4890 500}%
\special{pa 4834 542}%
\special{pa 4858 544}%
\special{pa 4866 566}%
\special{pa 4890 500}%
\special{fp}%
\put(47.8000,-8.6000){\makebox(0,0)[lb]{$2\lambda_{\varphi}^{(\varphi,c)}$}}%
%
\special{pn 13}%
\special{pa 4900 2670}%
\special{pa 4610 2260}%
\special{fp}%
%
\special{pn 13}%
\special{pa 4610 2270}%
\special{pa 4900 1860}%
\special{fp}%
%
\special{pn 13}%
\special{pa 4750 2080}%
\special{pa 4790 2010}%
\special{fp}%
\special{sh 1}%
\special{pa 4790 2010}%
\special{pa 4740 2058}%
\special{pa 4764 2056}%
\special{pa 4774 2078}%
\special{pa 4790 2010}%
\special{fp}%
%
\special{pn 13}%
\special{pa 4810 2530}%
\special{pa 4730 2440}%
\special{fp}%
\special{sh 1}%
\special{pa 4730 2440}%
\special{pa 4760 2504}%
\special{pa 4766 2480}%
\special{pa 4790 2478}%
\special{pa 4730 2440}%
\special{fp}%
%
\special{pn 13}%
\special{ar 4350 2270 256 256  0.0000000 6.2831853}%
%
\special{pn 13}%
\special{pa 4310 2020}%
\special{pa 4380 2020}%
\special{fp}%
\special{sh 1}%
\special{pa 4380 2020}%
\special{pa 4314 2000}%
\special{pa 4328 2020}%
\special{pa 4314 2040}%
\special{pa 4380 2020}%
\special{fp}%
%
\special{pn 13}%
\special{pa 4390 2530}%
\special{pa 4300 2530}%
\special{fp}%
\special{sh 1}%
\special{pa 4300 2530}%
\special{pa 4368 2550}%
\special{pa 4354 2530}%
\special{pa 4368 2510}%
\special{pa 4300 2530}%
\special{fp}%
\put(42.9000,-23.3000){\makebox(0,0)[lb]{$\phi$}}%
\put(36.9000,-23.2000){\makebox(0,0)[lb]{$\lambda'^{(\varphi)}$}}%
\put(47.2000,-23.2000){\makebox(0,0)[lb]{$\lambda'^{(c)}$}}%
%
\special{pn 8}%
\special{pa 410 300}%
\special{pa 260 400}%
\special{fp}%
%
\special{pn 8}%
\special{pa 270 390}%
\special{pa 270 1320}%
\special{fp}%
%
\special{pn 8}%
\special{pa 1830 2800}%
\special{pa 1950 2710}%
\special{fp}%
%
\special{pn 8}%
\special{pa 1940 2720}%
\special{pa 1940 1790}%
\special{fp}%
%
\special{pn 8}%
\special{pa 1940 1790}%
\special{pa 1780 1700}%
\special{fp}%
%
\special{pn 8}%
\special{pa 270 1310}%
\special{pa 420 1390}%
\special{fp}%
\put(22.2000,-8.7000){\makebox(0,0)[lb]{$+$}}%
\put(36.2000,-8.6000){\makebox(0,0)[lb]{$+$}}%
\put(21.2000,-23.1000){\makebox(0,0)[lb]{$+$}}%
\put(33.2000,-23.0000){\makebox(0,0)[lb]{$+$}}%
%
\special{pn 13}%
\special{ar 2820 2210 202 202  1.8925469 4.3757142}%
%
\special{pn 13}%
\special{ar 2850 2220 202 202  5.0398701 6.2831853}%
\special{ar 2850 2220 202 202  0.0000000 1.2356736}%
%
\special{pn 13}%
\special{pa 2750 2020}%
\special{pa 2910 2020}%
\special{dt 0.045}%
%
\special{pn 13}%
\special{pa 2740 2410}%
\special{pa 2900 2410}%
\special{dt 0.045}%
%
\special{pn 13}%
\special{pa 2910 2020}%
\special{pa 3250 1610}%
\special{fp}%
%
\special{pn 13}%
\special{pa 2740 2010}%
\special{pa 2430 1600}%
\special{fp}%
%
\special{pn 13}%
\special{pa 3230 2800}%
\special{pa 2920 2390}%
\special{fp}%
%
\special{pn 13}%
\special{pa 2400 2810}%
\special{pa 2740 2400}%
\special{fp}%
%
\special{pn 13}%
\special{pa 2530 2640}%
\special{pa 2600 2560}%
\special{dt 0.045}%
\special{sh 1}%
\special{pa 2600 2560}%
\special{pa 2542 2598}%
\special{pa 2566 2600}%
\special{pa 2572 2624}%
\special{pa 2600 2560}%
\special{fp}%
%
\special{pn 13}%
\special{pa 3110 2650}%
\special{pa 3040 2560}%
\special{dt 0.045}%
\special{sh 1}%
\special{pa 3040 2560}%
\special{pa 3066 2626}%
\special{pa 3074 2602}%
\special{pa 3098 2600}%
\special{pa 3040 2560}%
\special{fp}%
%
\special{pn 13}%
\special{pa 2640 1860}%
\special{pa 2570 1780}%
\special{dt 0.045}%
\special{sh 1}%
\special{pa 2570 1780}%
\special{pa 2600 1844}%
\special{pa 2606 1820}%
\special{pa 2630 1818}%
\special{pa 2570 1780}%
\special{fp}%
%
\special{pn 13}%
\special{pa 3020 1880}%
\special{pa 3110 1780}%
\special{dt 0.045}%
\special{sh 1}%
\special{pa 3110 1780}%
\special{pa 3052 1816}%
\special{pa 3074 1820}%
\special{pa 3080 1844}%
\special{pa 3110 1780}%
\special{fp}%
\put(26.6000,-18.0000){\makebox(0,0)[lb]{$2\lambda_{\varphi}^{(\varphi,c)}$}}%
\put(26.3000,-28.6000){\makebox(0,0)[lb]{$2\lambda_{\varphi}^{(\varphi,c)}$}}%
%
\special{pn 13}%
\special{pa 3050 2250}%
\special{pa 3050 2140}%
\special{fp}%
\special{sh 1}%
\special{pa 3050 2140}%
\special{pa 3030 2208}%
\special{pa 3050 2194}%
\special{pa 3070 2208}%
\special{pa 3050 2140}%
\special{fp}%
%
\special{pn 13}%
\special{pa 2620 2250}%
\special{pa 2620 2140}%
\special{fp}%
\special{sh 1}%
\special{pa 2620 2140}%
\special{pa 2600 2208}%
\special{pa 2620 2194}%
\special{pa 2640 2208}%
\special{pa 2620 2140}%
\special{fp}%
\put(33.1000,-8.5000){\makebox(0,0)[lb]{$\lambda_{\varphi}^{(c)}$}}%
\put(38.1000,-8.6000){\makebox(0,0)[lb]{$\lambda_{\varphi}^{(\varphi)}$}}%
%
\special{pn 13}%
\special{pa 3900 2530}%
\special{pa 3950 2440}%
\special{fp}%
\special{sh 1}%
\special{pa 3950 2440}%
\special{pa 3900 2490}%
\special{pa 3924 2488}%
\special{pa 3936 2508}%
\special{pa 3950 2440}%
\special{fp}%
\end{picture}%
\end{center}
\end{figure}
\end{center}
Here, the factor $2 \times$ stems from the fact that there are two ways
to contract external lines with non-local interaction points.
The contributions from the first and fourth diagrams are canceled 
for the case with (\ref{OSp-invL}).
Then we find that $\delta(\lambda_{\varphi}^{(\varphi)}v(x_1, x_2)) 
= \delta(\lambda_{\varphi}^{(c)}v(x_1, x_2))
= \delta(\lambda_{\varphi}^{(\varphi,c)}v(x_1, x_2))$,
when the relations (\ref{OSp-invL}) hold,
and an extra minus sign is not required for the propagation of $c_{\varphi}$
in case that the loop is not composed of $c_{\varphi}$ alone.

In this way, it is shown that,
if the relations (\ref{OSp-invL}) and the above feature relating $c_{\varphi}$
hold and quadratic divergences are removed,
the light field $\phi$ receives neither any quantum corrections from heavy fields $\varphi$ and $c_{\varphi}$
nor large corrections from the self-interaction,
$\varphi$ and $c_{\varphi}$ receive exactly the same size of radiative corrections,
and hence the mass hierarchy is stabilized against quantum corrections
at the one-loop level.

\section{Prototype model for grand unification}

Let us present a prototype model of a grand unified theory in a hidden form,
composed of massless fields including incomplete matter fields and ghost ones,
by reference to the model (C) in Sect. \ref{Symmetry reduction with ghost administration}.
The Lagrangian density possessing $SU(5)$ gauge invariance is given by,
\begin{eqnarray}
\hspace{-0.9cm}&~& \mathcal{L}_{\rm GUT} 
= \mathcal{L}_{\rm GUT}^{(1)} + \mathcal{L}_{\rm GUT}^{(2)}~,
\label{L-GUT}\\
\hspace{-0.9cm}&~& \mathcal{L}_{\rm GUT}^{(1)} =
- \frac{1}{4} F_{\mu\nu}^{\alpha} F^{\alpha \mu\nu}
+ (D_{\mu} H)^{\dagger} (D^{\mu} H) - \lambda_{H} (H^{\dagger} H)^2~,
\label{L-GUT1}\\
\hspace{-0.9cm}&~& \mathcal{L}_{\rm GUT}^{(2)}
= \sum_{k} {\Psi}_{{\rm L}k}^{\dagger} i \overline{\sigma}^{\mu} D_{\mu} \Psi_{{\rm L}k} 
+ \sum_{k'} {\Psi}_{{\rm R}k'}^{\dagger} i {\sigma}^{\mu} D_{\mu} \Psi_{{\rm R}k'} + \cdots~,
\label{L-GUT2}
\end{eqnarray}
where $D_{\mu} = \partial_{\mu} + i g_{\rm U} A_{\mu}^{\alpha} T^{\alpha}$,
$H=(H_C, H_W)^T$ is the Higgs boson with the fundamental representation $\bm{5}$,
and $\Psi_{{\rm L}k}$ and $\Psi_{{\rm R}k}$ are left-handed and right-handed Weyl fermions.
$A_{\mu}^{\alpha}$ $(\alpha = 1, \cdots, 24)$ are $SU(5)$ gauge bosons,
and $g_{\rm U}$ is the unified gauge coupling constant.
$H_C$ and $H_W$ stand for the colored components and the weak ones in $H$, respectively.
The ellipsis stands for terms such as Yukawa interactions.
We do not consider them, because it depends on the origin of matter fields

In the introduction of ghost fields $C_{\mu}$ and $c_{H_{C}}$, 
whose gauge quantum numbers are same as those of $X_{\mu}$ and $H_C$
((${\bm 3}, {\bm 2})$ and $({\bm 3}, {\bm 1})$
of $SU(3)_{\rm C} \times SU(2)_{\rm L}$, respectively),
the following Lagrangian density can be added,
\begin{eqnarray}
&~& \mathcal{L}_{\rm gh}^{(1)} = -(D'_{\mu} C_{\nu})^{\dagger}(D'^{\mu} C^{\nu}) 
+ (D'_{\mu} C_{\nu})^{\dagger}(D'^{\nu} C^{\mu})
+ (D'_{\mu} c_{H_{C}})^{\dagger} (D'^{\mu} c_{H_{C}})~,
\label{L-ghSU5(1)}\\
&~& \mathcal{L}_{\rm int}^{(1)}
= - \frac{ig_{\rm U}}{2} F_{\mu\nu}'^a \left(C^{\dagger\nu}C^{\mu} - C^{\dagger\mu}C^{\nu}\right)^a
\nonumber \\
&~& ~~~~~~~~~~~~~~ 
+ \frac{g_{\rm U}^2}{2}\left(X_{\nu}^{\dagger}X_{\mu} - X_{\mu}^{\dagger} X_{\nu}\right)^a \star
\left(C^{\dagger\nu} C^{\mu} - C^{\dagger\mu} C^{\nu}\right)^a
\nonumber \\
&~& ~~~~~~~~~~~~~~  
+ \frac{g_{\rm U}^2}{4} 
\left(C_{\nu}^{\dagger} C_{\mu} - C_{\mu}^{\dagger} C_{\nu}\right)^a
\star \left(C^{\dagger\nu}C^{\mu} - C^{\dagger\mu} C^{\nu}\right)^a
\nonumber \\
&~& ~~~~~~~~~~~~~~  
+ \frac{g_{\rm U}^2}{2} H_W^{\dagger} C_{\mu}^{\dagger} C^{\mu} H_W
- \frac{g_{\rm U}^2}{2} H_C^{\dagger} C_{\mu} C^{\dagger\mu} H_C
+ \frac{g_{\rm U}^2}{2} 
c_{H_C}^{\dagger} \left(X_{\mu} X^{\dagger\mu} - C_{\mu} C^{\dagger\mu}\right)c_{H_C}
\nonumber \\
&~& ~~~~~~~~~~~~~~  
+ \frac{ig_{\rm U}}{\sqrt{2}} (D'_{\mu} c_{H_C})^{\dagger} C^{\mu} H_W 
- \frac{ig_{\rm U}}{\sqrt{2}} H_W^{\dagger} C_{\mu}^{\dagger} (D'^{\mu} c_{H_C})
- \frac{ig_{\rm U}}{\sqrt{2}} (D'_{\mu} H_W)^{\dagger} C^{\dagger\mu} c_{H_C} 
\nonumber \\
&~& ~~~~~~~~~~~~~~  
+ \frac{ig_{\rm U}}{\sqrt{2}} c_{H_C}^{\dagger} C_{\mu} (D'^{\mu} H_W)
- 2 \lambda_H  \left({H_C}^{\dagger} H_C + {H_W}^{\dagger} H_W\right) 
\star c_{H_C}^{\dagger} c_{H_C}
\nonumber \\
&~& ~~~~~~~~~~~~~~  
- \lambda_H c_{H_C}^{\dagger} c_{H_C} \star c_{H_C}^{\dagger} c_{H_C}~, 
\label{L-intSU5(1)}
\end{eqnarray}
where $D'_{\mu} = \partial_{\mu} + i g_{\rm U} A_{\mu}^a T^a$, 
and $F_{\mu\nu}'^a$ is the field strength of the SM gauge bosons $A_{\mu}^a$
$(a=1,2 \cdots, 8, 21, \cdots, 24)$.
Using (\ref{L-GUT1}), (\ref{L-ghSU5(1)}) and (\ref{L-intSU5(1)}),
we obtain the relation
\begin{eqnarray}
\left. \mathcal{L}_{\rm light}^{(1)} =
\mathcal{L}_{{\rm GUT}\star}^{(1)} + \mathcal{L}_{\rm gh}^{(1)} + \mathcal{L}_{\rm int}^{(1)}
= \mathcal{L}_{{\rm SM}\star}^{(1)} 
+ {\bm \delta}_{\rm F} {\bm \delta}_{\rm F}^{\dagger} \Delta\mathcal{L}^{(1)}\right|_{M_{\rm U}}~,
\label{L-(1)}
\end{eqnarray}
where $\mathcal{L}_{{\rm GUT}\star}^{(1)}$ is the Lagrangian density
that the self-interactions of $X_{\mu}$ and $H$ are replaced by
the non-local ones in (\ref{L-GUT1}),
$\mathcal{L}_{{\rm SM}\star}^{(1)}$ and $\Delta\mathcal{L}^{(1)}$ are given by
\begin{eqnarray}
&~& \mathcal{L}_{{\rm SM}\star}^{(1)} =
- \frac{1}{4} F_{\mu\nu}'^{a} F'^{a \mu\nu}
+ (D'_{\mu} H_W)^{\dagger} (D'^{\mu} H_W) 
- \lambda_{H} H_W^{\dagger} H_W \star H_W^{\dagger} H_W~,
\label{L-SM(1)}\\
&~& \Delta\mathcal{L}^{(1)}
= - (D'_{\mu} X_{\nu})^{\dagger}(D'^{\mu} X^{\nu}) + (D'_{\mu} X_{\nu})^{\dagger}(D'^{\nu} X^{\mu})
- \frac{ig_{\rm U}}{2} F_{\mu\nu}'^{a} \left(X^{\dagger\nu}X^{\mu} - X^{\dagger\mu}X^{\nu}\right)^a
\nonumber \\
&~& ~~~~~~~~~~~~~~ 
+ \frac{g_{\rm U}^2}{8}\left(X_{\nu}^{\dagger} X_{\mu} - X_{\mu}^{\dagger} X_{\nu}\right)^a \star
\left(X^{\dagger\nu}X^{\mu} - X^{\dagger\mu}X^{\nu}\right)^a
\nonumber \\
&~& ~~~~~~~~~~~~~~ 
- \frac{g_{\rm U}^2}{8}\left(C_{\nu}^{\dagger} C_{\mu} - C_{\mu}^{\dagger} C_{\nu}\right)^a \star
\left(C^{\dagger\nu}C^{\mu} - C^{\dagger\mu}C^{\nu}\right)^a
\nonumber \\
&~& ~~~~~~~~~~~~~~ 
 + (D'_{\mu} {H_C})^{\dagger} (D'^{\mu} H_C)
+ \frac{g_{\rm U}^2}{4} 
\left(H_C^{\dagger} X_{\mu} X^{\dagger\mu} H_C  
+ c_{H_C}^{\dagger} C_{\mu} C^{\dagger\mu} c_{H_C}\right)
\nonumber \\
&~& ~~~~~~~~~~~~~~ 
+ \frac{g_{\rm U}^2}{2} H_W^{\dagger} X_{\mu}^{\dagger} X^{\mu} H_W
+ \frac{ig_{\rm U}}{\sqrt{2}}(D'_{\mu} {H_C})^{\dagger} X^{\mu} H_W 
- \frac{ig_{\rm U}}{\sqrt{2}} H_W^{\dagger} X_{\mu}^{\dagger} (D'^{\mu} H_C)
\nonumber \\
&~& ~~~~~~~~~~~~~~ 
- \frac{ig_{\rm U}}{\sqrt{2}} (D'_{\mu} H_W)^{\dagger} X^{\dagger\mu} H_C 
+ \frac{ig_{\rm U}}{\sqrt{2}} {H_C}^{\dagger} X_{\mu} (D'^{\mu} H_W)
\nonumber \\
&~& ~~~~~~~~~~~~~~ 
- \frac{\lambda_{H}}{2} \left({H_C}^{\dagger} H_C \star {H_C}^{\dagger} H_C
- c_{H_C}^{\dagger} c_{H_C} \star c_{H_C}^{\dagger} c_{H_C}\right)
- 2 \lambda_{H} {H_C}^{\dagger} H_C \star {H_W}^{\dagger} H_W~.
\label{DeltaL(1)}
\end{eqnarray}
Note that the self-interaction of the weak Higgs boson $H_W$ is given as the non-local one
in (\ref{L-SM(1)}), but it is regarded as the local one 
if the fundamental length $\ell$ is small enough.

The $\mathcal{L}_{\rm light}^{(1)}$ is invariant under the fermionic transformations,
\begin{eqnarray}
&~& {\bm \delta}_{\rm F} H_C = - c_{H_C}~,~~{\bm \delta}_{\rm F} H_C^{\dagger} = 0~,~~ 
{\bm \delta}_{\rm F} c_{H_C} = 0~,~~ {\bm \delta}_{\rm F} c_{H_C}^{\dagger} = H_C^{\dagger}~,~~
{\bm \delta}_{\rm F} X_{\mu} = - C_{\mu}~,~~
{\bm \delta}_{\rm F} X_{\mu}^{\dagger} = 0~,~~
\nonumber \\ 
&~& {\bm \delta}_{\rm F} C_{\mu} = 0~,~~ 
{\bm \delta}_{\rm F} C_{\mu}^{\dagger} = X_{\mu}^{\dagger}~,~~
{\bm \delta}_{\rm F} H_W = 0~,~~
{\bm \delta}_{\rm F} H_W^{\dagger} = 0~,~~
{\bm \delta}_{\rm F} A_{\mu}^{a} = 0
\label{delta-F-HW}
\end{eqnarray}
and
\begin{eqnarray}
&~& {\bm \delta}_{\rm F}^{\dagger} H_C = 0~,~~
{\bm \delta}_{\rm F}^{\dagger} H_C^{\dagger} =  c_{H_C}^{\dagger}~,~~ 
{\bm \delta}_{\rm F}^{\dagger} c_{H_C} = H_C~,~~
{\bm \delta}_{\rm F}^{\dagger} c_{H_C}^{\dagger} = 0~,~~
{\bm \delta}_{\rm F}^{\dagger} X_{\mu} = 0~,~~
{\bm \delta}_{\rm F}^{\dagger} X_{\mu}^{\dagger} = C_{\mu}^{\dagger}~,~~
\nonumber \\ 
&~& {\bm \delta}_{\rm F}^{\dagger} C_{\mu} = X_{\mu}~,~~ 
{\bm \delta}_{\rm F}^{\dagger} C_{\mu}^{\dagger} = 0~,~~
{\bm \delta}_{\rm F}^{\dagger} H_W = 0~,~~
{\bm \delta}_{\rm F}^{\dagger} H_W^{\dagger} = 0~,~~
{\bm \delta}_{\rm F}^{\dagger} A_{\mu}^{a} = 0~.
\label{delta-F-HW-dagger}
\end{eqnarray}

Next, we consider the matter part $\mathcal{L}_{\rm GUT}^{(2)}$,
taking three types of matter multiplets such as 
$\Psi_{\overline{\bm 5}_{\rm L}}$, $\Psi_{{\bm 5}_{\rm R}}$
and $\Psi_{{\bm{10}}_{\rm L}}$.\\
(a) Starting from $\Psi_{\overline{\bm 5}_{\rm L}} = (d_{\rm L}^c, l_{\rm L})^T$, 
after introducing the ghost partner $c_{l_{\rm L}}$ of $l_{\rm L}$,
the kinetic term of down type $SU(2)_{\rm L}$-singlet quark $d_{\rm L}^c$ is derived as follows,
\begin{eqnarray}
&~& {\Psi}_{\overline{\bm 5}_{\rm L}}^{\dagger} i \overline{\sigma}^{\mu} 
D_{\mu} \Psi_{\overline{\bm 5}_{\rm L}} 
+ c_{l_{\rm L}}^\dagger i \overline{\sigma}^{\mu} D'_{\mu} c_{l_{\rm L}}
- \frac{g_{\rm U}}{\sqrt{2}} d_{\rm L}^{c\dagger} \overline{\sigma}^{\mu} C_{\mu}^{*} c_{l_{\rm L}}
+ \frac{g_{\rm U}}{\sqrt{2}} c_{l_{\rm L}}^{\dagger} \overline{\sigma}^{\mu} C_{\mu}^{T} d_{\rm L}^c
\nonumber \\
&~& ~~ \left. = d_{\rm L}^{c\dagger} i \overline{\sigma}^{\mu} D'_{\mu} d_{\rm L}^c 
+ {\bm \delta}_{\rm F}  {\bm \delta}_{\rm F}^{\dagger}
\left({l_{\rm L}}^\dagger i \overline{\sigma}^{\mu} D'_{\mu} l_{\rm L}
+ \frac{g_{\rm U}}{\sqrt{2}} d_{\rm L}^{c\dagger} \overline{\sigma}^{\mu} X_{\mu}^{*} l_{\rm L}
- \frac{g_{\rm U}}{\sqrt{2}} {l_{\rm L}}^{\dagger} \overline{\sigma}^{\mu} X_{\mu}^{T} d_{\rm L}^c\right)
\right|_{M_{\rm U}}~,
\label{Rd}
\end{eqnarray}
where the superscripts $c$ and $*$ represent the operation of charge conjugation
and complex conjugation, respectively,
and the fermionic transformations are given by
\begin{eqnarray}
&~& {\bm \delta}_{\rm F} l_{\rm L} = -c_{l_{\rm L}}~,~~
{\bm \delta}_{\rm F} l_{\rm L}^{\dagger} = 0~,~~ 
{\bm \delta}_{\rm F} c_{l_{\rm L}} = 0~,~~ 
{\bm \delta}_{\rm F} c_{l_{\rm L}}^{\dagger} = -l_{\rm L}^{\dagger}~,~~
{\bm \delta}_{\rm F} X_{\mu}^T = - C_{\mu}^T~,~~
{\bm \delta}_{\rm F} X_{\mu}^{*} = 0~,~~
\nonumber \\
&~&  
{\bm \delta}_{\rm F} C_{\mu}^T = 0~,~~ {\bm \delta}_{\rm F} C_{\mu}^{*} = X_{\mu}^{*}~,~~
{\bm \delta}_{\rm F} d_{\rm L}^c = 0~,~~
{\bm \delta}_{\rm F} d_{\rm L}^{c\dagger} = 0
\label{delta-F-dLc}
\end{eqnarray}
and
\begin{eqnarray}
&~& {\bm \delta}_{\rm F}^{\dagger} l_{\rm L} = 0~,~~
{\bm \delta}_{\rm F}^{\dagger} l_{\rm L}^{\dagger} = -c_{l_{\rm L}}^{\dagger}~,~~ 
{\bm \delta}_{\rm F}^{\dagger} c_{l_{\rm L}} = l_{\rm L}~,~~ 
{\bm \delta}_{\rm F}^{\dagger} c_{l_{\rm L}}^{\dagger} = 0~,~~
{\bm \delta}_{\rm F}^{\dagger} X_{\mu}^T = 0~,~~
{\bm \delta}_{\rm F}^{\dagger} X_{\mu}^{*} = C_{\mu}^{*}~,~~
\nonumber \\
&~&  
{\bm \delta}_{\rm F}^{\dagger} C_{\mu}^T = X_{\mu}^{T}~,~~ 
{\bm \delta}_{\rm F}^{\dagger} C_{\mu}^{*} = 0~,~~
{\bm \delta}_{\rm F}^{\dagger} d_{\rm L}^c = 0~,~~
{\bm \delta}_{\rm F}^{\dagger} d_{\rm L}^{c\dagger} = 0~.
\label{delta-F-dLc-dagger}
\end{eqnarray}\\
(b) From the content with $\Psi_{{\bm 5}_{\rm R}} = (d_{\rm R}, l_{\rm R}^c)^T$ 
and the ghost partner $c_{d_{\rm R}}$ of $d_{\rm R}$,
the kinetic term of $SU(2)_{\rm L}$-doublet lepton $l_{\rm L} = (l_{\rm R}^c)^c$ is derived as
(after the charge conjugation is performed),
\begin{eqnarray}
&~& {\Psi}_{{\bm 5}_{\rm R}}^{\dagger} i {\sigma}^{\mu} D_{\mu} \Psi_{{\bm 5}_{\rm R}} 
+ c_{d_{\rm R}}^\dagger i {\sigma}^{\mu} D'_{\mu} c_{d_{\rm R}}
+ \frac{g_{\rm U}}{\sqrt{2}} c_{d_{\rm R}}^{\dagger} {\sigma}^{\mu} C_{\mu} l_{\rm R}^c
- \frac{g_{\rm U}}{\sqrt{2}} l_{\rm R}^{c\dagger} {\sigma}^{\mu} C_{\mu}^{\dagger} c_{d_{\rm R}}
\nonumber \\
&~& ~~ \left. = l_{\rm R}^{c\dagger} i {\sigma}^{\mu} D'_{\mu} l_{\rm R}^c 
+ {\bm \delta}_{\rm F}  {\bm \delta}_{\rm F}^{\dagger}
\left({d_{\rm R}}^\dagger i {\sigma}^{\mu} D'_{\mu} d_{\rm R}
 - \frac{g_{\rm U}}{\sqrt{2}} {d_{\rm R}}^{\dagger} {\sigma}^{\mu} X_{\mu} l_{\rm R}^c
 + \frac{g_{\rm U}}{\sqrt{2}} l_{\rm R}^{c\dagger} {\sigma}^{\mu} X_{\mu}^{\dagger} d_{\rm R}\right)
\right|_{M_{\rm U}}~,
\label{Rl}
\end{eqnarray}
where the fermionic transformations are given by
\begin{eqnarray}
{\bm \delta}_{\rm F} d_{\rm R} = - c_{d_{\rm R}}~,~~
{\bm \delta}_{\rm F} d_{\rm R}^{\dagger} = 0~,~~ 
{\bm \delta}_{\rm F} c_{d_{\rm R}} = 0~,~~ 
{\bm \delta}_{\rm F} c_{d_{\rm R}}^{\dagger} = -d_{\rm R}^{\dagger}~,~~
{\bm \delta}_{\rm F} l_{\rm R}^c = 0~,~~ 
{\bm \delta}_{\rm F} l_{\rm R}^{c\dagger} = 0
\label{delta-F-lRc}
\end{eqnarray}
and
\begin{eqnarray}
{\bm \delta}_{\rm F}^{\dagger} d_{\rm R} = 0~,~~
{\bm \delta}_{\rm F}^{\dagger} d_{\rm R}^{\dagger} = -c_{d_{\rm R}}^{\dagger}~,~~ 
{\bm \delta}_{\rm F}^{\dagger} c_{d_{\rm R}} = d_{\rm R}~,~~ 
{\bm \delta}_{\rm F}^{\dagger} c_{d_{\rm R}}^{\dagger} =0~,~~
{\bm \delta}_{\rm F}^{\dagger} l_{\rm R}^c = 0~,~~ 
{\bm \delta}_{\rm F}^{\dagger} l_{\rm R}^{c\dagger} = 0~.
\label{delta-F-lRc-dagger}
\end{eqnarray}\\
(c) From the content with $\Psi_{{\bm{10}}_{\rm L}} = \{q_{\rm L}, u_{\rm L}^c, e_{\rm L}^c\}$ 
and the ghost partners $c_{u_{\rm L}^c}$ of $u_{\rm L}^c$ 
and $c_{e_{\rm L}^c}$ of $e_{\rm L}^c$,
the kinetic term of $SU(2)_{\rm L}$-doublet quark $q_{\rm L}$ is derived as
\begin{eqnarray}
&~& {\Psi}_{{\bm{10}}_{\rm L}}^{\dagger} 
i \overline{\sigma}^{\mu} D_{\mu} \Psi_{{\bm{10}}_{\rm L}} 
+ c_{u_{\rm L}^c}^\dagger i \overline{\sigma}^{\mu} D'_{\mu} c_{u_{\rm L}^c}
+ c_{e_{\rm L}^c}^\dagger i \overline{\sigma}^{\mu} D'_{\mu} c_{e_{\rm L}^c}
\nonumber \\
&~& 
+ \frac{g_{\rm U}}{\sqrt{2}} c_{u_{\rm L}^c}^{\dagger} \overline{\sigma}^{\mu} C_{\mu} q_{\rm L}
- \frac{g_{\rm U}}{\sqrt{2}} q_{\rm L}^{\dagger} \overline{\sigma}^{\mu} C_{\mu}^{\dagger} c_{u_{\rm L}^c}
+ \frac{g_{\rm U}}{\sqrt{2}} q_{\rm L}^{\dagger} \overline{\sigma}^{\mu} C_{\mu} c_{e_{\rm L}^c}
- \frac{g_{\rm U}}{\sqrt{2}} c_{e_{\rm L}^c}^{\dagger} \overline{\sigma}^{\mu} C_{\mu}^{\dagger} q_{\rm L}
\nonumber \\
&~&~~  = q_{\rm L}^{\dagger} i \overline{\sigma}^{\mu} D'_{\mu} q_{\rm L} 
+ {\bm \delta}_{\rm F} {\bm \delta}_{\rm F}^{\dagger}
\Biggl({u_{\rm L}^c}^\dagger i \overline{\sigma}^{\mu} D'_{\mu} u_{\rm L}^c
+ {e_{\rm R}}^\dagger i {\sigma}^{\mu} D'_{\mu} e_{\rm R}
\nonumber \\
&~&~~~~ \left. \left.
- \frac{g_{\rm U}}{\sqrt{2}} {u_{\rm L}^c}^{\dagger} \overline{\sigma}^{\mu} X_{\mu} q_{\rm L} 
+ \frac{g_{\rm U}}{\sqrt{2}} q_{\rm L}^{\dagger} \overline{\sigma}^{\mu} X_{\mu}^{\dagger} u_{\rm L}^c
- \frac{g_{\rm U}}{\sqrt{2}} q_{\rm L}^{\dagger} \overline{\sigma}^{\mu} X_{\mu} e_{\rm L}^c
+ \frac{g_{\rm U}}{\sqrt{2}} 
{e_{\rm L}^c}^{\dagger} \overline{\sigma}^{\mu} X_{\mu}^{\dagger} q_{\rm L}\right)
\right|_{M_{\rm U}}~,
\label{Rq}
\end{eqnarray}
where the fermionic transformations are given by
\begin{eqnarray}
&~& {\bm \delta}_{\rm F} u_{\rm L}^c = -c_{u_{\rm L}^c}~,~~
{\bm \delta}_{\rm F} u_{\rm L}^{c\dagger} = 0~,~~ 
{\bm \delta}_{\rm F} c_{u_{\rm L}^c} = 0~,~~ 
{\bm \delta}_{\rm F} c_{u_{\rm L}^c}^{\dagger} = -u_{\rm L}^{c\dagger}~,~~
\nonumber \\
&~& {\bm \delta}_{\rm F} e_{\rm L}^{c\dagger} = -c_{e_{\rm L}^c}^{\dagger}~,~~
{\bm \delta}_{\rm F} e_{\rm L}^{c} = 0~,~~ 
{\bm \delta}_{\rm F} c_{e_{\rm L}^c}^{\dagger} = 0~,~~ 
{\bm \delta}_{\rm F} c_{e_{\rm L}^c} = -e_{\rm L}^{c}~,~~
{\bm \delta}_{\rm F} q_{\rm L} = 0~,~~
{\bm \delta}_{\rm F} q_{\rm L}^{\dagger} = 0
\label{delta-F-qL}
\end{eqnarray}
and
\begin{eqnarray}
&~& {\bm \delta}_{\rm F}^{\dagger} u_{\rm L}^c = 0~,~~
{\bm \delta}_{\rm F}^{\dagger} u_{\rm L}^{c\dagger} = -c_{u_{\rm L}^c}^{\dagger}~,~~ 
{\bm \delta}_{\rm F}^{\dagger} c_{u_{\rm L}^c} = u_{\rm L}^{c}~,~~ 
{\bm \delta}_{\rm F}^{\dagger} c_{u_{\rm L}^c}^{\dagger} = 0~,~~
\nonumber \\
&~& {\bm \delta}_{\rm F}^{\dagger} e_{\rm L}^{c\dagger} = 0~,~~
{\bm \delta}_{\rm F}^{\dagger} e_{\rm L}^{c} = -c_{e_{\rm L}^{c}}~,~~ 
{\bm \delta}_{\rm F}^{\dagger} c_{e_{\rm L}^c}^{\dagger} = e_{\rm L}^{c\dagger}~,~~ 
{\bm \delta}_{\rm F}^{\dagger} c_{e_{\rm L}^c} = 0~,~~
{\bm \delta}_{\rm F}^{\dagger} q_{\rm L} = 0~,~~
{\bm \delta}_{\rm F}^{\dagger} q_{\rm L}^{\dagger} = 0~.
\label{delta-F-qL-dagger}
\end{eqnarray}

The kinetic terms of $u_{\rm L}^c$ and $e_{\rm L}^c$ should be added,
by introducing $u_{\rm L}^c$ and $e_{\rm L}^c$ as $\Phi'_{\rm o}$ (ordinary fields belonging to
multiplets of the SM gauge group).

This model has following excellent features.
The unification of the SM gauge coupling constants 
occurs such that $g_3 = g_2 = g_1 = g_{\rm U}$ at $M_{\rm U}$.
The triplet-doublet splitting of Higgs boson is realized
in the form that $H_W$ becomes the Higgs doublet in the SM and 
the ghost $c_{H_C}$ makes $H_C$ unphysical.
The longevity of proton is fully guaranteed
because both $X$ gauge bosons and their ghost partners are unphysical.

\end{document}